\documentclass[sn-mathphys]{sn-jnl}  
\usepackage{txfonts}
\usepackage{natbib}
\usepackage{graphicx}
\usepackage{tabularx}
\usepackage[pagewise,switch]{lineno}
\usepackage{mathtools}
\usepackage[utf8]{inputenc}
\usepackage[caption=false]{subfig}
\usepackage{physics}
\jyear{2021}
\usepackage{xcolor}

\theoremstyle{thmstyleone}%
\theoremstyle{thmstyletwo}%

\theoremstyle{thmstylethree}%

\begin{document}

\title[PRATUSH experiment concept and design overview]{PRATUSH experiment concept and design overview}

\author*[1]{\fnm{Mayuri} \sur{Sathyanarayana Rao}}\email{mayuris@rri.res.in}
\author[1]{\fnm{Saurabh} \sur{Singh}}\email{saurabhs@rri.res.in}
\author[1]{\fnm{Srivani} \sur{K.S.}}\email{vani\_4s@rri.res.in}
\author[1]{\fnm{Girish} \sur{B.S.}}\email{bsgiri@rri.res.in}
\author[1]{\fnm{Keerthipriya} \sur{Satish}}\email{keerthi@rri.res.in}
\author[1]{\fnm{Somashekar} \sur{R.}}\email{som@rri.res.in}
\author[1]{\fnm{Raghunathan} \sur{Agaram}}\email{raghu@rri.res.in}
\author[1]{\fnm{Kavitha} \sur{Kalyanasundaram}}\email{kavitha@rri.res.in}
\author[1]{\fnm{Gautam} \sur{Vishwapriya}}\email{vishwapriya@rri.res.in}
\author[1]{\fnm{Ashish} \sur{Anand}}\email{ashishanand@rri.res.in}
\author[1]{\fnm{Udaya Shankar} \sur{N.}}\email{uday@rri.res.in}
\author[1]{\fnm{Seetha} \sur{S.}}\email{seetha@rri.res.in}

\affil*[1]{ \orgname{Raman Research Institute}, \orgaddress{\street{C V Raman Avenue, Sadashivanagar}, \city{Bengaluru}, \postcode{560080}, \state{Karnataka}, \country{India}}}


\abstract{PRATUSH -- Probing ReionizATion of the Universe using Signal from Hydrogen -- is a proposed  cosmology experiment to detect the global red-shifted 21-cm signal from the Cosmic Dawn and Epoch of Reionization (CD/EoR). PRATUSH orbiting the Moon will seek to precisely measure the low-frequency radio sky-spectrum over 40 to 200 MHz. The scientific observations would be made in the radio-quiet region when in the farside of the Moon, and the data would be transmitted back to Earth when in the near-side. PRATUSH was proposed to the Indian Space Research Organization (ISRO) during a call for proposals in the announcement of opportunity for science payloads in 2018. PRATUSH is in the pre-project studies phase. Here we present a mission concept and baseline design of the proposed payload optimized to operate over the Cosmic Dawn signal band of 55 - 110 MHz. Starting with a description of the fundamental design principles followed, we discuss the PRATUSH baseline design and sensitivity. We further enumerate the challenges that are common to most PRATUSH like experiments, which have been proposed to seek a detection of the CD/EoR signal in orbit in the lunar farside. Due to the highly sensitive nature of the measurement, PRATUSH is designed to operate as a solo experiment with a dedicated spacecraft. Our simulations, assuming a mission lifetime of two years, estimate that PRATUSH would have the sensitivity required to detect the CD signal predicted by the standard models with varying degrees of confidence.A concept model of PRATUSH is under development, which is expected to lead to the engineering model followed by flight model subject to mission approval.}


\maketitle
\section{Introduction}
Post the epoch of recombination ($z\sim 1100$) most of the baryonic matter in the Universe comprised neutral hydrogen and helium and was devoid of any sources of radiation. This is referred to as the Dark Ages \citep{1977SvAL....3..155V,2000PhR...333..203R}. Following the Dark Ages, the first sources of radiation (such as stars and galaxies) formed in a period which is aptly called the Cosmic Dawn (CD) \citep{2010PhRvD..82b3006P}. The radiation from these first sources eventually re-ionized the Universe in the Epoch of Reionization (EoR) \citep{2006PhR...433..181F}. Very little is known about CD and EoR. From previous studies, it is estimated that CD and EoR spanned redshifts $15 \lesssim\ z\  \lesssim20$ and $6 \lesssim\ z\  \lesssim15$ respectively, though the precise redshifts, the nature of the first sources, and the physics of the re-ionization process is poorly understood \citep{1965ApJ...142.1633G,2006ARA&A..44..415F}.
The abundant hydrogen content of the Universe can be used as a probe of CD and EoR \citep{2012RPPh...75h6901P}. Specifically, the evolution of the average strength of the spin-flip transition of neutral hydrogen, having a rest-wavelength of 21-cm, can be studied as a function of cosmic time through CD and EoR \citep{1979MNRAS.188..791H,1999A&A...345..380S}.  This signal appears as a weak monopole-, or global-signal, throughout the sky \citep{2008PhRvD..78j3511P}. The exact strength and spectral shape of the signal is determined by the astrophysics and cosmology of CD and EoR, and the redshift of these processes determines the observed frequency. A comprehensive review of the science and probes of reionization is presented in \cite{Zaroubi2013}. A variety of plausible global-21cm signals predicted by standard physics is presented in \cite{Cohen2017}. The parameter space of CD and EoR is wide open \citep{10.1093/mnras/stab2737}. A high-confidence signal detection remains elusive. PRATUSH -- Probing ReionizATion of the Universe using Signal from Hydrogen -- seeks to do just that.

\section{Motivation}
There exist several mature experiments seeking to make a confirmed detection of the global signal from CD and EoR from Earth. Some of these are - SARAS : the Shaped Antenna measurement of the background RAdio Spectrum \citep{SARAS2_inst}, EDGES : Experiment to Detect the Global Epoch-of-Reionization Signature \citep{EDGES2018}, CTP : Cosmic Twilight Polarimeter \citep{Nhan2019}, REACH : Radio Experiment for the Analysis of Cosmic Hydrogen \citep{REACH2019}, and LEDA : Large-Aperture Experiment to Detect the Dark Age \citep{Bernardi2018}. Several exciting results have emerged from ground-based experiments. For instance, SARAS2 \citep{SARAS2_results} was the first experiment to place constraints on the astrophysical nature of the first sources and rule out late heating models. EDGES presented a detection of a wide and deep absorption at 78 MHz \citep{EDGES2018}. If this was indeed a true cosmological signal detection, attributed to the CD then it would contradict standard models of cosmology and requires invoking non-standard physics, such as radiatively interacting dark matter \citep{Barkana2018,Fialkov2018,Munoz2018} or a mechanism for energy injection in the early Universe which results in an excess radio-background \citep{Fialkov2019,Ewall-Wice2018,Feng2018}. However, a more straightforward explanation for this surprising measurement is confusion arising from artefacts introduced by the experimental setup. It has been suggested that by varying the foreground separation methods and parameters, there exist sinusoidal oscillations in the EDGES data set which are degenerate with deep and wide absorption trough suggested by the EDGES team \citep{tauscher2020, Hills2018, SSRS2019}. It is conceivable that such oscillations in the data could be an artefact from standing waves in cables or a frequency-dependent beam \citep{Sims2020}, or artefacts from the ground plane \citep{Bradley2019}, which are all simple and more likely explanations not invoking exotic physics. The SARAS-3 experiment \citep{SARAS3} has  experimentally ruled out the cosmological origin of the reported EDGES detection to $95\%$ confidence, thus restoring faith in standard models. The true signal from CD remains undetected, due to the inherently faint nature of the signal, instrument and environment related challenges as listed above.
Furthermore, terrestrial Radio Frequency Interference (RFI) poses a major challenge to ground-based experiments. In particular the FM radio band of $88-108$ MHz lies entirely within the frequency space of CD experiments (typically centered around 80 MHz). RFI may be observed by experiments either via direct line-of-sight pick-up or reflected off surrounding terrain, including ``moonshine" reflecting RFI off the moon \citep{vedantham2014} and space-debris \citep{tingay2013}. While most site selection surveys are quick to identify strong RFI in the form of spectral lines, low-lying RFI and broadband emission are difficult to detect prior to scientific observation deployments.
In addition to trace residual RFI present in terrestrial radio quiet regions, CD/EoR detection experiments on Earth are undoubtedly subject to the refractive and attenuation effects of the ionosphere. It remains uncertain if ionospheric effects are entirely detrimental to Earth based global CD/EoR signal detection experiments \citep{Shen2021,Datta2016,Sokolowski2015}.  Recent work has shown that the topographical effects of objects in the horizon of the antenna beam on Earth are non-trivial \citep{Bassett2021}. In addition to objects in the horizon, the inhomogeneous and stratified nature of the soil column beneath an antenna on Earth, albeit below a reflector, can potentially appear as artefacts in a sensitive, high dynamic range experiment seeking to detect the cosmological signal amidst bright foregrounds. This has necessitated experiments to move from deploying over land to over waterbodies \citep{JN2021}.  Any experiment on Earth is more sensitive to these near field effects of `coupling to' soil or water   than one in free space.
Control of instrument generated systematics, a clean measurement environment, and robust data-analysis methods, all play a vital role in validating the cosmological origin of any signal detection claim.  
Thus a high-confidence signal detection requires measurement in an environment that produces the least confusing artefacts and robust analysis techniques for data interpretation. 
Measurements from the Radio Astronomy Explorer 2 (RAE-2) have suggested that the Moon is an attenuator to terrestrial RFI over the frequency range of 25 kHz to 13 MHz \citep{Alexander1975}. Simulations suggest that the Moon attenuates terrestrial RFI over the frequency range of CD and EoR detection experiments \citep{Bassett2020}.  The atmosphere of the Moon comprises gases with an infinitesimally small density compared to Earth. These factors, namely the attenuation of terrestrial RFI and a very thin atmosphere, make the lunar farside an ideal site to detect the signal from the formation of the first stars and galaxies. Thus an orbiter observing in the lunar farside provides conditions ideal to overcome systematics routinely faced on Earth.

Unsurprisingly, several experiments have been proposed to study the low-frequency radio sky in this pristine environment. The pioneering experiment proposal, which was the first to propose and lay the ground work and make the case for a lunar farside CD/EoR experiment was DARE : Dark Ages Reionization Explorer \citep{burns2017}. We acknowledge the DARE team for their contributions that have flagged off the race to the lunar farside for CD/EoR signal detection. These include DAPPER : Dark Ages Polarimeter PathfindER  \citep{DAPPER}, LuSEE : Lunar Surface Electromagnetics Experiment \citep{LUSEE2020}, ROLSES : Radio wave Observations at the Lunar Surface of the photoElectron Sheath \citep{ROLSES} and FARSIDE: Farside Array for Radio Science Investigations of the Dark ages and Exoplanets (FARSIDE) \citep{FARSIDE} are some of the proposed and upcoming experiments seeking to observe the low-frequency radio sky from the lunar farside either in orbit or with landers. 

Of note is the Netherlands-China Low Frequency Explorer (NCLE) \citep{NCLE2021}, which is a scientific payload onboard the Chinese satellite Queqiao. Using three 5 metre monopole antennas and electronic receivers, NCLE seeks to study the radio sky over a frequency range of 80 kHz to 80 MHz, with a host of technical and scientific objectives. The spacecraft orbits the Earth-Moon L2 point at a height of $\sim 64000$ km over the lunar farside. Following instrument commissioning tests, the three monopole antennas were deployed in  November 2019 and NCLE started making scientific measurements, the published results of which are awaited. 


PRATUSH  -- Probing ReionizATion of the Universe using Signal from Hydrogen --  from India joins this league of experiments. PRATUSH was proposed to the Indian Space Research Organization (ISRO) during a call for proposals in the announcement of opportunity for science payloads in 2018. As of the writing of this paper, PRATUSH is funded in the pre-project studies phase. PRATUSH will operate in lunar orbit making scientific measurements when in the lunar farside in an environment that is expected to be devoid of terrestrial RFI. PRATUSH is unique in its design as it will be purpose built to have a smooth calibrated bandpass spectrum, leveraging the difference in spectra of foregrounds and the cosmological signal over sufficiently wide bandwidths. This combination of radio quiet location coupled with system design based on smooth bandpass will set PRATUSH apart from several ground-based and space-based counterparts attempting to detect the cosmological signal from CD and EoR. The details of the payload instrument design philosophy are presented in Section \ref{sec:design}.


\section{Scientific objectives}
PRATUSH is a custom experiment with a primary science goal of detecting the global 21-cm signal from CD and EoR. To eliminate any electromagnetic interference from other payloads, PRATUSH is envisioned as the sole experiment on a dedicated space-craft. The system design and mission concept are optimized for the primary science goal. We describe here the primary science goal and the secondary science that PRATUSH data products will naturally enable. 

\subsection{Primary Science Objective: Detection of the red-shifted global 21-cm signal from the CD and EoR}
\begin{figure*}
    \centering
    \includegraphics[width=\textwidth]{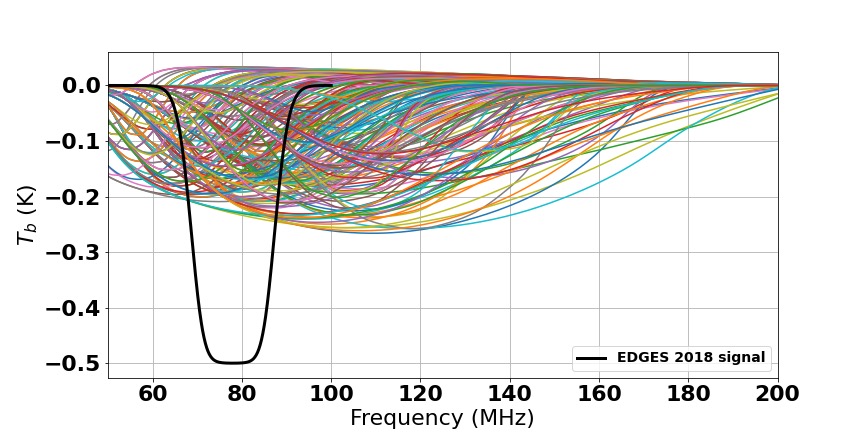}
    \caption{The thin colored lines depict an atlas of the shapes and strengths of global CD and EoR signals as predicted by standard physics as derived from the models in \citep{Cohen2017}. The solid black line is the signal detection claimed by the EDGES experiment \citep{EDGES2018}. The parameters space of CD and EoR remains wide open.}
    \label{fig:Eor_models}
\end{figure*}

The $\Lambda$CDM model of cosmology posits that following the end of the epoch of recombination (redshift $z\sim1100$), baryonic matter in the Universe was mostly neutral. Several observations have confirmed that the late Universe is predominantly ionized. The cosmological periods over which the first sources of radiation formed and whose radiation subsequently re-ionized the Universe are termed the Cosmic Dawn (CD) and Epoch of Reionization (EoR) respectively ($80\gtrsim z \gtrsim 6$). The 21-cm signal, arising from the spin-flip transition of hydrogen atom, is one of the most promising observational probes of CD and EoR. It is predicted that the variation of spatially-averaged  intensity of this signal with frequency -- expected to span 40-200~MHz after cosmological redshift -- can be observed as a global signature whose spectral shape and strength can be used to directly infer the astrophysics and cosmology of CD and EoR ~\citep{Shaver1999,Cohen2017}. The contrast of brightness temperature of the redshifted $21-$cm signal against the background radiation temperature is the quantity of interest, given by equation \ref{eq:Tb},

\begin{equation}\label{eq:Tb}
        \delta T_b = 27 x_{HI}(1+\delta_b)\Big(\frac{\Omega_bh^2}{0.023}\Big)\Big(\frac{0.15}{\Omega_mh^2}\frac{1+z}{10}\Big)^{\frac{1}{2}} \times\Big(\frac{T_s-T_R}{T_s}\Big)\Big[\frac{\partial_r v_r}{(1+z)H(z)}\Big]\ \textrm{mK}
\end{equation}
, wherein $x_{HI}$ is the neutral hydrogen fraction, $\Omega_b$ is the baryon density, $\Omega_m$ is the matter density, $T_s$ is the spin temperature, $T_r$ is the radiation temperature, $\delta_b$ is the fractional overdensity in baryons. $h$ is the reduced Hubble constant, and the last term is from the line of sight velocity gradient. Thus, this differential brightness temperature $ \delta T_b$  encompasses cosmological and astrophysical quantities and their average temporal variation over CD/EoR. The spin temperature varies with redshift and is coupled either to the radiation temperature or gas temperature mediated via Lyman-alpha photons \citep{Wouthuysen:1952}. Additionally there are subtle effects resulting from Lyman alpha and radiation heating \citep{Reis2021}. All these processes leave an imprint in the specific shape of the global signal from CD and EoR. Figure~\ref{fig:Eor_models} represents a range of plausible signals posited by standard physics and cosmology \citep{Cohen2017}. Also shown is the reported signal detection by EDGES. Recent results from SARAS-3 have rejected the cosmological origins of this signal with 95.3\% confidence \citep{SARAS3}, however, the true CD/EoR signal has not yet been detected with high confidence.

By accurately measuring the low-frequency radio sky spectrum over 40-200 MHz in the clean measurement environment provided in orbit over the lunar farside, PRATUSH will detect the true signal from CD/EoR using a purpose built receiver. 

\subsection{Secondary Science: An accurate measurement of the radio-sky spectrum over 40 - 200 MHz}
A precise measurement of the low-frequency radio sky is essential for effective separation between cosmological and astrophysical signals. Precise measurements from ground-based CD/EoR  detection experiments have provided corrections to the offset of all-sky radio maps \citep{Patra2015,Monsalve2021} and have measured  the temperature spectral index for different sky directions \citep{EDGES2008,EDGES2019}. This has provided insight into physical mechanisms such as the steepening of the Galactic synchrotron spectrum by an aging electron population and spectral flattening due to the nature of intrinsic injection spectrum. These measurements, though an improvement over older ones, continue to be limited by the same effects that plague ground-based 21-cm cosmology, namely systematics via coupling of the antenna to the Earth beneath, terrestrial RFI, and ionosphere. By observing in space in the lunar farside PRATUSH will improve our understanding of the radiophysics of galactic synchrotron emission and diffuse gas in the interstellar medium. In the age of mega telescopes, such as the Square Kilometre Array (SKA), PRATUSH will serve to provide the cleanest absolute calibration for sky maps. This is especially relevant as the all-sky maps that form the basis for most global sky models are several decades old \citep{landecker1970,haslam1982}, with inherent calibration errors that range from $\sim 1 - \sim 10\%$ \citep{GSM,MSR2017a}. The effect of these errors is further exacerbated when interpolating between these sky maps generated from different telescopes. 

\subsection{Secondary Science: Study of low-frequency radio environment around the Moon }
In the post space-race era, there is renewed interest in lunar studies from space agencies around the world. From habitability to planet studies, the Moon is the closest and most accessible extra-terrestrial astronomical object. There are a range of experiments proposed to be in the lunar farside, either in orbit or deployed with landers. Several ambitious experiments propose radio interferometers over large baselines to observe the low-frequency radio sky from the vantage point of the lunar farside. The lunar farside is a desirable site due to the expectation of low to no terrestrial radio frequency interference. However, the measurements of the level radio quietness around the Moon over the frequency range 25 kHz to 13 MHz by Radio Astronomy Explorer 2~ (RAE-2) are almost 50 years old \citep{Alexander1975}. While simulations show that the Moon can provide adequate shielding to most terrestrial RFI \citep{Bassett2020}, these do not account for effects of RFI from electronics onboard satellites and orbiters around the Moon. The expensive and ambitious efforts of deploying large experiments in the lunar farside to achieve a range of science goals will be limited in scope if the quality of the radio environment is compromised by RFI from satellites and their electronics. The PRATUSH dataset by its very nature will be sensitive to and measure the RFI spectrum over its full frequency of operation when observing around the Moon. With a mission lifetime of 2 Earth years and an expected orbital period of 2 Earth hours, PRATUSH will record the lunar farside radio environment over several epochs of observations. 


\section{Scientific payload design principles}\label{sec:design}
Physically motivated simulations, to the extent modeled using existing all sky maps, show that foreground spectra are devoid of inflections if observed by an ideal instrument. This feature can be used to separate foregrounds from the CD/EoR signal which has several inflections or turning points over sufficiently wide bandwidths. It is these turning points which in their shape and strength that encode meaningful astrophysical and cosmological parameters \citep{2012MNRAS.424.2551M,2013ApJ...777..118M,2014Natur.506..197F}. Thus, essential to a high confidence detection of the CD/EoR signal is the requirement that there must be no instrument generated or data-analysis induced shapes, by means of errors in bandpass calibration or systematics such as `additive contaminants' that potentially confuse a signal detection. This motivates the design of PRATUSH to produce a bandpass calibrated and antenna return-loss corrected spectrum that is smooth and thermal noise dominated (or devoid of systematic shapes) to levels below the expected signal strength. This criterion drives the sub-system level and system level requirements. We employ Maximally Smooth (MS) functions \citep{MSR2015} as the primary metric to define smoothness for foreground separation and for characterising systematics levels in the instrument design. MS functions are expected to minimally subsume the signal over wide-enough bandwidths that include sufficient signal turning points. This aids in distinguishing predicted CD/EoR  signals from foregrounds that are expected to be smooth, devoid of inflections or turning points \citep{MSR2017b}. 

We extend the application of MS functions to quantify the instrument bandpass. We require that the time-averaged, bandlimited, bandpass calibrated spectrum acquired on terminating the receiver with standard terminations (open, short, and $50 \Omega$ load) is described by an MS function with the resulting residual being nearly Gaussian with a Root Mean Square (RMS) value of $\sim$ milliKelvin or below. This ensures that the receiver is thermal noise dominated to levels below typically expected MS baseline subtracted CD/EoR signal strength, that is, there are no systematics in the bandpass calibrated receiver response that can confuse or dominate CD/EoR signal detection. This criteria governs our choice of receiver architecture and calibration methodology. The general design practices that aid in the final goal of a smooth calibrated instrument response are detailed in \ref{sec:length} and \ref{sec:EMI_philosophy}. We place separate requirements to qualify the antenna behavior, including its return loss and beam chromaticity, and once again use MS functions for the same. This is described further in \ref{section:Antenna_philosophy} and \ref{section:Antenna_design}.

\subsection{Antenna beam and return loss}\label{section:Antenna_philosophy}
A wideband antenna is the primary sensing element of PRATUSH. It converts the radio sky-brightness to voltage at its terminals, which is then amplified and measured by the receiver electronics. The sensitivity of the antenna as a function of direction (in azimuth and elevation) is determined by the antenna beam shape. The impedance matching of the antenna to receiver electronics as defined by a standard 50$\Omega$ reference, i.e the antenna return loss, defines the coupling of the voltage at the antenna feed point to the electronics downstream. The return loss of the antenna determines the efficiency with which the antenna can radiate power input at the terminals (or reciprocally transfer the detected power to the receiver). This along with resistive losses (such as by heating) determines the total efficiency of the antenna. The shape of the observed sky spectrum, which contains the cosmological signal, is modified by the shape of the antenna's return loss, beam pattern, and efficiency. The design specification for the PRATUSH antenna is that it must not introduce any spectral structure in the observed sky-spectrum that could potentially dominate or confuse the signature expected from the CD/EoR  signal. For this, PRATUSH ideally requires an antenna that has a frequency-independent beam and a smooth return-loss over the full frequency of observation (that is, return loss that can be described by an MS function). The next best thing is to have an antenna whose return loss and beam minimally confuses signal detection, as described further in \ref{section:Antenna_design}. Wideband frequency-independent antennas adopting a similar guideline have been designed and used in the SARAS series of experiments from the same team \citep{Wideband_antennas,SARAS3_antenna}.

\subsection{Short cable lengths}\label{sec:length}
Standing waves in cables interconnecting receiver electronics that are uncalibrated by the bandpass calibration mechanism can result in artefacts that confuse a global CD/EoR signal detection. Standing waves originating from signals reflected internal to the receiver due to impedance mismatches, such as between an amplifier and a coaxial cable, are particularly detrimental. They are additive to the sky-signal and not multiplicative. Thus they are not effectively calibrated or removed by traditional bandpass calibration methods. In an effort to minimize such standing waves emphasis is placed on high quality impedance matching between cascaded devices and minimizing cable lengths connecting them. The period of a standing wave of an electromagnetic frequency supported by a cable is inversely proportional to its length. As a rule of thumb the periodicity in the frequency structure of a standing wave $\delta f$ over a cable of length $L$ is given by equation \ref{eq:standing}.
\begin{equation}\label{eq:standing}
    \delta f = \frac{v_{light}}{2\times \sqrt{\epsilon_r}\times L}
\end{equation}
where $\frac{v_{light}}{\sqrt{\epsilon_r}}$ gives the velocity of light in the medium. Rapidly oscillating standing waves are challenging to separate from the slowly varying CD/EoR signal of unknown exact form. Thus, the CD/EoR signal is best distinguished from a standing wave of periodicity that is large enough be a flat, smooth, or DC-like signal over the bandwidth. Assuming the velocity factor of copper to be 0.95, the natural standing wave period for a copper cable of length $1~m$ is $\sim 142.5$ MHz. Since the CD/EoR  signal has turning points spaced over the range of $\sim 10$ MHz,  all coaxial cables connecting components and modules in PRATUSH are no longer than $\sim$centimetres, such that standing waves are over a hundred times broader than the signal of interest. 

\subsection{Multi-level EMI shielding of electronics}\label{sec:EMI_philosophy}
Self-generated RFI, which we will refer to as Electromagnetic Inteference (EMI) henceforth,  constitutes radiation generated by satellite and payload electronics within the frequency range of observations, which contaminate the scientific dataset either in localized frequency channels or via broadband leakage. The strongest source of EMI is expected to be the digital receiver electronics of the payload and clocking electronics of the satellite package. In addition to channel loss from flagging of channels where clock lines reside, strong sources of EMI also introduce shapes by spectral leakage in the wings of the line, and via increase in the noise floor of the spectrum. The advantages of observing in a radio-quiet environment in the lunar farside are lost from unmanaged and unmitigated EMI. A multi-pronged approached is required to minimize and overcome EMI. Some strategies to mitigate these effects are adopting multilayer EMI shielding for the chassis of the payload and satellite electronics and operating the receiver over sub-octave continuous bandwidths to avoid effects of harmonics.

\section{PRATUSH system design}
While the goal design of PRATUSH targets the full CD/EoR band of 40-200 MHz, we choose the CD band over 55-110 MHz for the baseline design of PRATUSH described here. The primary challenge in moving from baseline to goal design is design of a multi-octave antenna that satisfies the antenna validation pipeline (see Section.\ref{sec:pipeline}). The design of a multi-octave antenna is challenging as is, the presence of the satellite bus in the near-field of the antenna and the added requirement of spectral smoothness in the antenna behaviour is a complicating factor. With this in mind, we ask the question about which octave band yield maximum scientific returns. This is quantified by (a) the band where the astrophysics is most poorly constrained by other measurements (b) presence of sufficient signal turning points thus increasing the ability to distinguish from foregrounds and (c) the bands most contaminated by radio frequency interference and benefit the most from lunar farside observations.
In addition to the redshifted 21-cm signal, Lyman-alpha is a powerful probe to study EoR \citep{2006ARA&A..44..415F,10.1111/j.1365-2966.2004.07594.x}. However, at earlier redshifts, the paucity of sources over the Cosmic Dawn makes Lyman-alpha a less feasible probe. Furthermore, various CD/EoR  experiments including EDGES \citep{2019ApJ...875...67M} and SARAS-2 \citep{SARAS2_results} have placed constraints on EoR, and have already ruled out models that suggest rapid reionization and inefficient X-ray heating. Thus, there is much to be gained from focusing on the less constrained period of CD. The frequency range of 55-110 MHz of global 21-cm signal which corresponds to CD, contains multiple turning points in the signal for a variety of standard models, which aids in the clean separation from smooth foregrounds thus maximizing the signal-to-noise ratio, where noise in this context includes foregrounds. Finally, the octave range of 55-110 MHz also encompasses the FM band of 88-110 MHz which is the most pernicious source of terrestrial RFI that plagues all ground-based experiments, and hence would be the ideal frequency range to observe in the radio quiet region afforded by the lunar farside.
Thus we choose $55-110$ MHz as the frequency range for the baseline design of PRATUSH.
The PRATUSH system comprises three main sub-systems namely:	
\begin{itemize}
    \item An antenna (comprising a monocone over a shaped reflector), which is the sensing element that transduces the electromagnetic field of radio-sky to voltage across its leads
    \item A Radio Frequency (RF) system that in turn comprises the analog receiver and a vector network analyser (VNA) section. The RF system  detects voltage signal from the antenna and performs the functions of bandpass calibration and signal conditioning (such as amplification and band-shaping), and in-situ antenna return loss measurement.
    \item A digital receiver that samples the analog signal, generates cross-correlation spectra, flags and rejects outlier channels in the spectra, averages data as per pre-programmed requirements, and appropriately stores the final spectra for downlinking to ground stations. 
\end{itemize}

\begin{figure}[h]
    \centering
    \begin{center}
            \includegraphics[clip=True, trim={4cm 0cm 0cm 0cm} , scale=0.3, width=\textwidth]{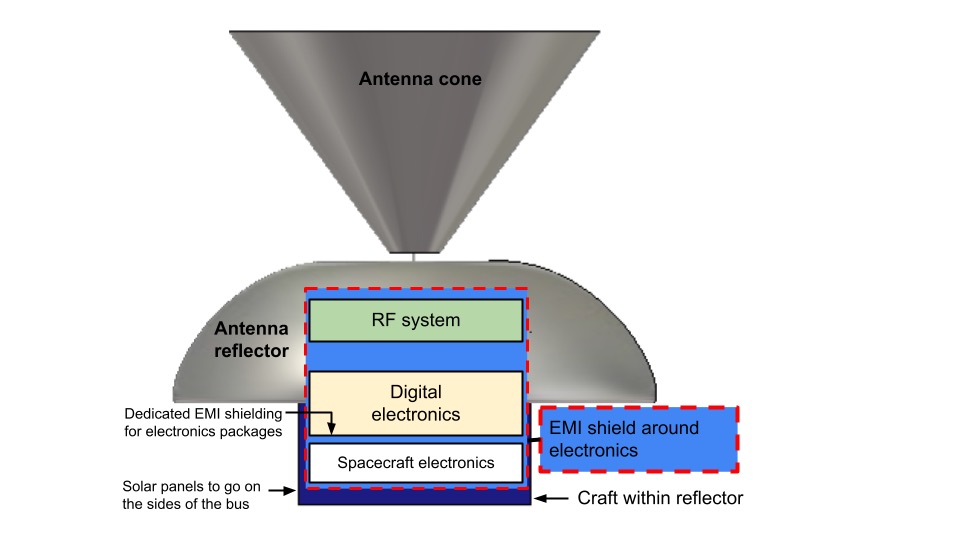}
    \caption{The PRATUSH system design. The antenna sits over the bus, with the craft being completely covered by the reflector of the antenna. The electronics, namely the RF system and digital receiver of the payload, as well as the spacecraft electronics are all well enclosed in one or several electromagnetic interference (EMI) shielding enclosures within the bus. Antennas used for communication and data transfer are not shown here.}
    \end{center}
    \label{fig:system_design}
\end{figure}

The instrument is designed such that the antenna sits over the bus, with the craft being partially covered by the reflector of the antenna. The electronics, namely the RF system and digital receiver of the experiment, as well as the spacecraft electronics are all well enclosed in one or several EMI shielding enclosures within the spacecraft, as seen in Figure~\ref{fig:system_design}. To minimize the EMI generated by electronics, and for the properties of the antenna to be well constrained, it is essential that PRATUSH has a dedicated craft and flies as an independent mission or jettisons from a mother ship and is injected into  the identified lunar orbit.

It is expected that PRATUSH will orbit the Moon in a low inclination orbit, having an average orbital period of $\sim2$ hours. We define the prime cone region that is most suitable for scientific observation as the region in the lunar farside in the shadow of the Earth and the Sun. Subject to the epoch of launch and insertion, multiple orbits have been explored with the final choice of orbit chosen to maximize the time spent in the prime cone region with minimal required orbital corrections (maximum orbit stability). The total usable time for science observations is determined by multiple factors including stabilization of electronics after entering the prime cone region. With a conservative estimate of 15\% of the total observing time being scientifically useful, and that we are able to model foregrounds in the bandpass calibrated average spectrum using MS functions (thus preserving CD signal amplitude) we estimate requiring a total of 200 hours in the prime-cone region over a mission life of 2 Earth years. The details of the orbit are beyond the scope of this paper.

\subsection{PRATUSH antenna} \label{section:Antenna_design}
 Designing a frequency-independent antenna with smooth return loss over wide-frequencies as required by PRATUSH is a challenge. While designing a multi-octave frequency-independent antenna is challenging as it were, the presence of a bus in the near-field of an antenna adds further complexity. The electro-mechanical model of the bus is very much a part of the antenna design wherein the  bus dimensions are determined by the payload and spacecraft electronics.
 Two parallel design approaches were explored to arrive at a baseline antenna design for PRATUSH. The first approach assumed a custom bus design and the second assumed standard bus-dimensions (such as cubesat dimensions). While the former allows for greater customization and flexibility, it also increases the parameter space for optimization and complexity in design. The latter method provides a starting framework over which the antenna must be designed to operate with no flexibility in bus dimension. An acceptable candidate antenna design for either case would have to meet the design criteria over a minimum frequency range of one octave within the CD/EoR signal band as described in Section \ref{sec:pipeline}. We optimize the antenna over the baseline design frequency range of $55-110$ MHz. We also note here that while 55-110 MHz is the minimum viable frequency range specification for the PRATUSH baseline design and consequently the antenna, the rest of the electronics, in principle, can be tuned to operate over much wider bandwidths by appropriate choice of filters. 

\begin{figure}
     \centering
 \resizebox{\hsize}{!}{
         \includegraphics{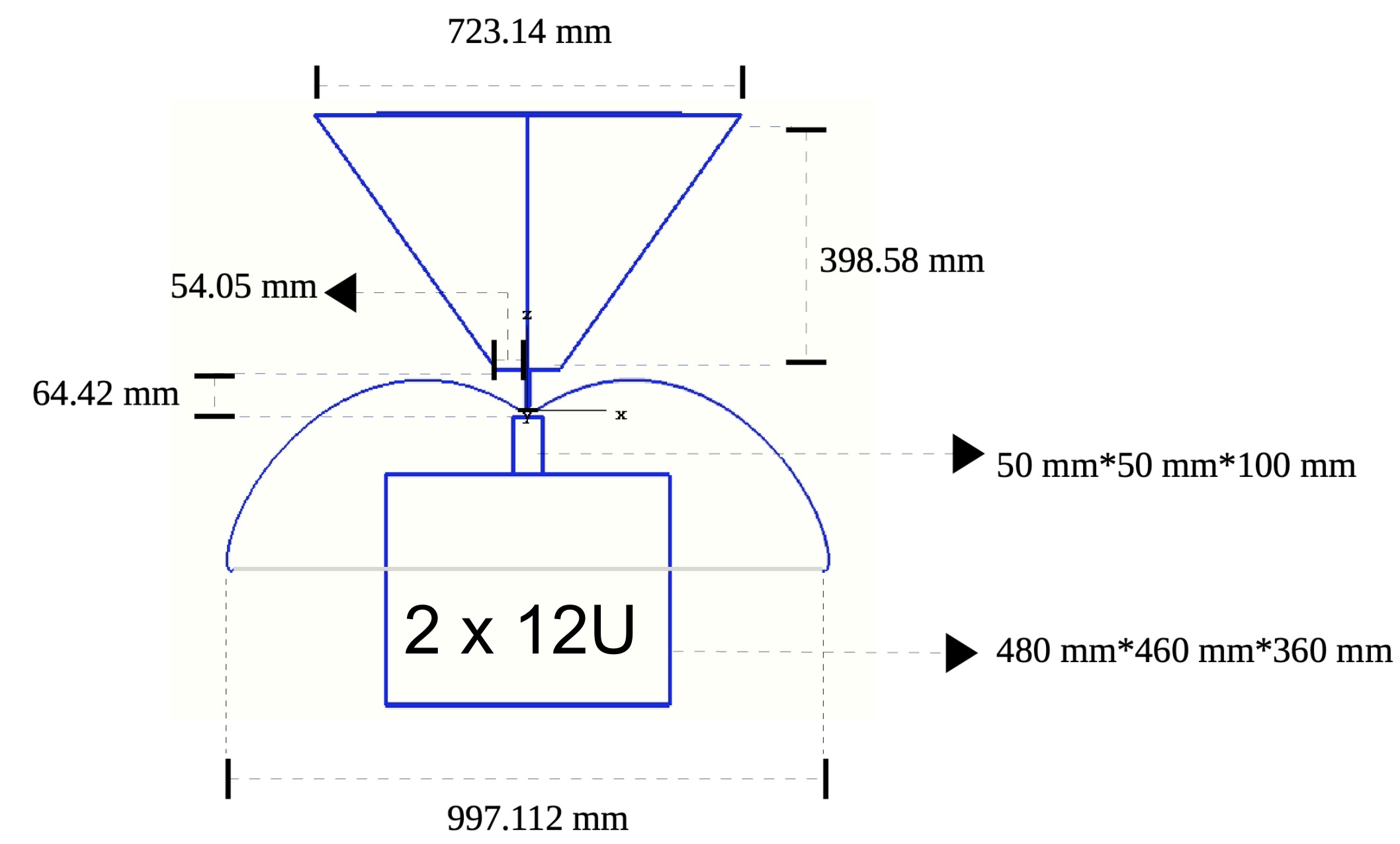}}
         \caption{Dimensions of the baseline design PRATUSH antenna shown in cross section}
         \label{fig:antenna_dim}
\end{figure}

\begin{figure}
         \centering
          \resizebox{\hsize}{!}{\includegraphics{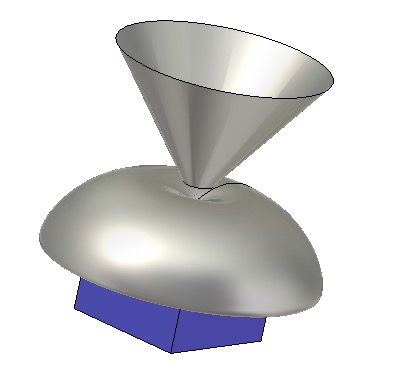}}
         \caption{An artist view of the simulated PRATUSH  antenna over a bus of dimension 2x12U (in blue). The radiator cone of the antenna and the reflector (shown in grey) are exposed outside the bus. The bus houses the payload electronics and the satellite electronics include telemetry, control, and power. The side faces of the bus are mounted with solar panels which are irradiated when outside the solar shadow region.}
         \label{fig:antenna_3D}
\end{figure}

\begin{figure*}
         \centering
         {\includegraphics[width=0.8\textwidth]{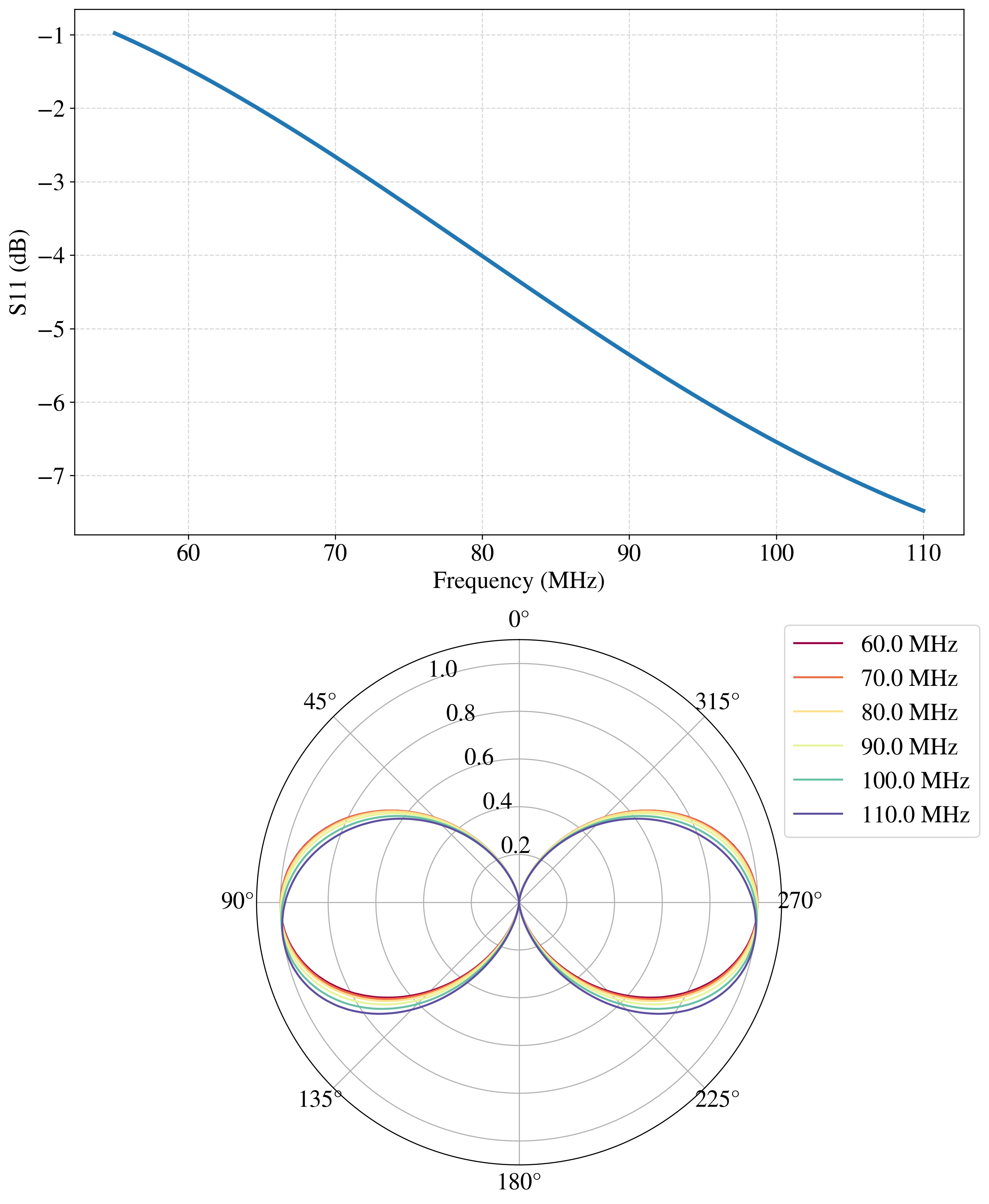}}\label{fig:antenna_specs}
         \caption{(a) The return loss of the antenna over 55-110 MHz. (b)  Cross section of the antenna beam as a function of elevation at azimuth=$0^{0}$. The beam is azimuthally symmetric.}         
\end{figure*}

The receiver bandpass is calibrated by a noise injection scheme as described in section \ref{sec:bpcal}, but the frequency dependent behaviour of the antenna is not calibrated by the same calibration mechanism. Thus particular attention is paid to the antenna design by enforcing criteria at multiple levels starting from EM simulation during the design phase. The antenna is qualified as acceptable for further testing and prototype fabrication once it passes a design validation pipeline. Based on this criterion a baseline design for PRATUSH antenna is a monocone antenna with a shaped reflector over a bus of dimension 2 x 12 U (corresponding to a dimension of 480 mm x 460 mm x 360 mm). Figure~\ref{fig:antenna_3D} shows a simulated view of the PRATUSH antenna over a 2x12U bus.  The shape of the reflector follows an optimzed log-spiral curve, to ensure impedance matching for a smooth return loss over 55-100 MHz as shown in Figure~\ref{fig:antenna_specs}. The cross section of the beam as a function of elevation and frequency shown in Figure~\ref{fig:antenna_specs}. 

\subsubsection{Antenna design validation pipeline}\label{sec:pipeline}
The two important specifications of the ideal PRATUSH antenna are a frequency-independent beam and a smooth return-loss over the design bandwidth. In practise,  we evaluate the effect of the antenna performance on the sky spectrum it would observe and consequently the cosmological signal it will observe, instead of placing thresholds to individually quantify the acceptable deviations from frequency independence and return loss smoothness. The inputs to the antenna validation pipeline are a sky model that is physically motivated, the simulated antenna beam pattern and return-loss over the frequency range of the simulation, a model for the lunar emissions that the antenna might receive via any backlobes and predicted models of the global 21-cm signal (such as those in Fig. \ref{fig:Eor_models}). For the physically motivated sky model we use GMOSS \citep{MSR2017a}. A simulated or mock sky-spectrum is generated by convolving the antenna beam with the sky model. The effective power and hence antenna temperature that enters the PRATUSH receiver is determined by the antenna return loss $\Gamma$ as given in equation \ref{eq:Gamma}.

\begin{equation}
    T_{A} = T_{sky}(1-\abs{\Gamma}^{2})
\end{equation}\label{eq:Gamma}

\begin{figure}[h]
    \centering
    \includegraphics[scale=0.45,angle =90]{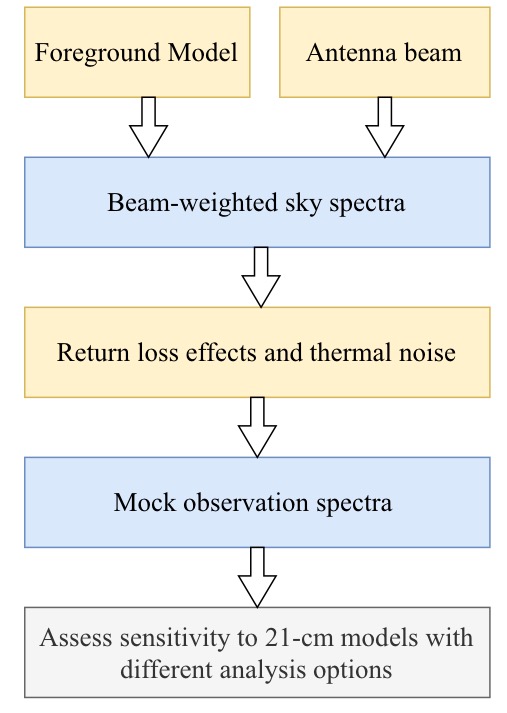}
    \caption{A flowchart representation of the PRATUSH antenna design qualification pipeline. The inputs to the pipeline are a physically motivated sky model, CD/EoR signal models from theoretical predictions and antenna parameters, namely the beam and return loss as a function of frequency. A MS function fit is subtracted from the resulting mock spectrum. The sensitivity of the antenna to signal detection is quantified by the difference in the RMS of the residual in the case when the signal is present in the pipeline input compared to when it is absent (null hypothesis).}
    \label{fig:pipeline_flowchart}
\end{figure}

The null hypothesis considers that there is no cosmological signal from CD/EoR present in the data, wherein $T_{sky} = T_{foreground}$, where $T_{foreground}$ includes the sky model weighted by the antenna primary beam and lunar emission. In the case where the signal from CD/EoR is present, $T_{sky} = T_{foreground} + T_{CD/EoR}$.
The resulting mock spectrum is then modeled with a MS function, which quantifies spectral smoothness and is expected to completely describe the foreground spectrum and minimally subsume the CD/EoR signal. The residual on subtracting the MS model from the mock sky-spectrum is used to qualify the antenna design as sensitive or otherwise to detect different CD/EoR signal models.
This is given by the ability to distinguish, with varying degrees of confidence, the residual RMS to fitting with an MS function, between the null hypothesis case and     the case when a signal is present. The pipeline is visually represented as flowchart in Figure~\ref{fig:pipeline_flowchart}  

\begin{figure}[h]
    \centering
    \includegraphics[width=\textwidth]{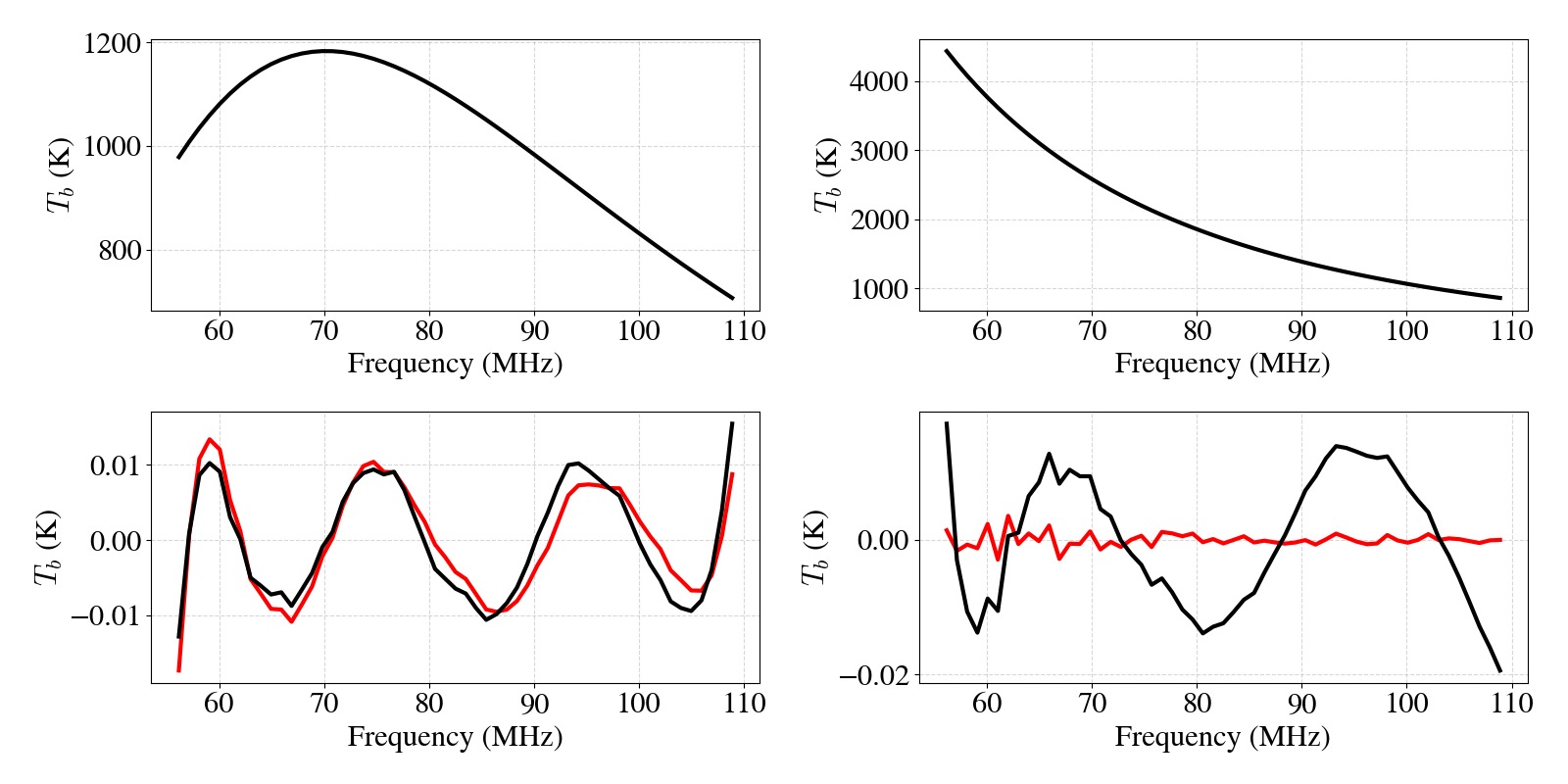}
    \caption{Validation of the baseline design PRATUSH antenna. No antenna correction is applied in the left panels, whereas in the right panels the antenna return loss is corrected for. Top left: The sky spectrum including modified by the antenna beam response and return loss. Top right : The sky spectrum after correcting for the antenna return loss. Bottom left : The residual on subtracting an MS function from the return-loss affected sky-spectrum for the cases when the CD signal is absent (red curve) and present (black curve) in the total spectrum. Bottom right : The residual on fitting the return-loss corrected intrinsic sky spectrum with an MS function for the case when the CD signal is absent (red curve) and present (black curve). Thus by correcting for the antenna return loss, we are able to distinguish between the presence and absence of the CD signal and the antenna design is qualified by the design validation pipeline.}
    \label{fig:antenna_validation}
\end{figure}

Several candidate antenna designs and iterations of these were tested by the validation pipeline of which the baseline PRATUSH antenna design presented in this paper emerged as the most suitable candidate.
A spectrum containing only the foreground as observed with the PRATUSH antenna can be adequately described using a MS function such that the RMS of the residual is $7.1$ milli-Kelvin. This is insufficient to distinguish between the case when the CD/EoR signal is present in the sky-spectrum from when it is absent, as seen in the left panel of Figure~\ref{fig:antenna_validation}. However, correcting for the antenna return loss, enabled by in-situ measurement (described in Section.\ref{sec:VNA}), we are able to make a clean distinction between the presence and absence of the CD signal in the mock sky-spectrum. This is demonstrated by levels of the residuals presented by the red and black curves in the right panel of Figure~\ref{fig:antenna_validation}.


\subsection{PRATUSH RF system}
\begin{figure*}[ht]
    \centering
    \includegraphics[scale=0.35]{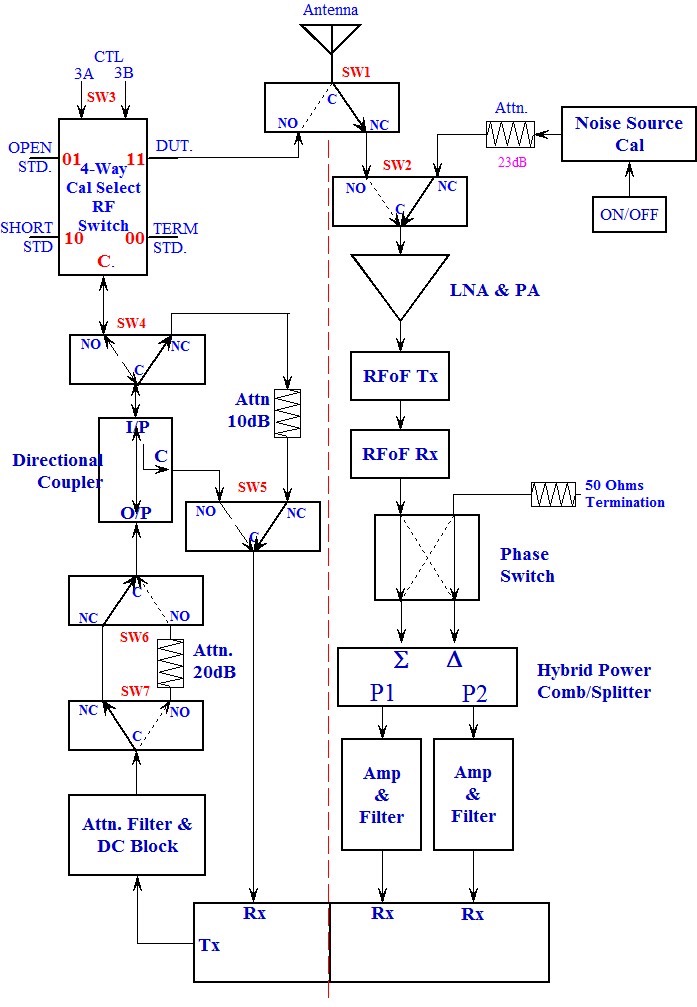}
    \caption{A block diagram representation of the PRATUSH RF system. Bandpass calibration is achieved in the analog receiver section using a double differencing technique using a Dicke switched noise source at the antenna and an RF switch further down the signal processing chain. A custom 'light' VNA section provides in-situ measurement of the antenna return loss properties, once each cycle to measure and correct for broad changes introduced in the antenna properties resulting from thermal cycling.}
    \label{fig:block_dia}
\end{figure*}

The PRATUSH RF system comprises the analog receiver and the Vector Network Analyzer (VNA) section. The analog receiver is responsible for bandpass calibration and signal conditioning. The VNA section measures the antenna return loss. The top-level RF system architecture block diagram is presented in Figure~\ref{fig:block_dia}.

The calibration scheme for PRATUSH is adapted from SARAS-3 \citep{JN2021}, with modifications made specific to PRATUSH. In PRATUSH the antenna is connected to a `master switch' that connects the antenna to either the analog receiver or to a VNA section. In the former switch position, the resulting configuration is in observation mode where as the latter configuration is in antenna characterization mode.

\subsubsection{PRATUSH receiver bandpass calibration} \label{sec:bpcal}
The receiver bandpass calibration is accomplished by injecting flat spectrum "cold" and "hot" noise signals, that are generated by a cascade of a matched attenuator and a noise source that form the bandpass calibrator unit. The received signal, after amplification, is split into two paths, the relative phase between which alternates between 0 and 180 degrees as part of calibration. These two signal paths are subsequently sampled and digitised by two separate ADCs and correlated in an FPGA, in the digital electronics. 
In all, the bandpass calibration involves cycling through 6 states, enabled by three switches or controls. These are detailed in Table \ref{tab:RF_analog_calibration}.
The states or consequently the spectra acquired when the receiver is connected to the sky via the antenna are termed $OBSxx$ where $xx$ can denote $00$ or $11$ depending on the phase switch position. States or consequently the spectra acquired when the receiver is connected to the attenuated noise source are denoted $CALxx$ where $xx$ can denote $00,01,10,\textrm{or }11$ depending on the position of the phase switch and the noise source being ON or OFF. The phase switch controls the phase of the signal as it is split into two arms as they enter the sum and difference ports of a $180$ degree hybrid. This phase switching enables subtraction of any low level noise that may be added by the digital circuitry, assuming they stay constant between the switching states. The Dicke switch, switches the analog receiver between the signal from the antenna and attenuated noise source.  The noise-source itself can be turned ON or OFF. In each calibration cycle, the spectra measured with this cold noise injection (noise-source OFF) are subtracted from antenna state spectra to remove the LNA noise spectrum to the first order.  

A bandpass calibrated spectrum for each calibration cycle is derived from spectra acquired from various states in that cycle as shown in equation \ref{eq:measurement_equation}. 

\begin{equation}\label{eq:measurement_equation}
    T_{bp} = T_{abs}\frac{(OBS00 - OBS11) - (CAL00 - CAL01) }{(CAL10 - CAL11)-(CAL00 - CAL01)}
\end{equation}

\begin{table}[h]
    \centering
     \resizebox{\textwidth}{!}{
    \begin{tabular}{ccccc}
         Master (SW1) & Dicke switch (SW2) & Noise source & Phase switch & State\\
         \hline 
         OFF & 0 & OFF & 0 & OBS00\\
         OFF & 0 & OFF & 1 & OBS11\\
         OFF & 1 & OFF & 0 & CAL00\\
         OFF & 1 & OFF & 1 & CAL01\\
         OFF & 1 & ON & 0 & CAL10\\
         OFF & 1 & ON & 1 & CAL11\\
    \end{tabular}
    }
    \caption{The bandpass calibration cycle of the analog receiver in the PRATUSH RF system. The master switch connects the antenna to the analog receiver, and the Dicke switch, Noise source power and phase switch are toggled to cycle the receiver over the various states. The spectra measured in all these states are then used for the receiver bandpass calibration.}
    \label{tab:RF_analog_calibration}
\end{table} 
 , wherein $T_{abs}$ is a constant obtained by the absolute calibration of the receiver using physical hot and cold loads of known absolute temperature. In order to isolate the sensitive electronics close to the antenna from any noise injected by the digital receiver, PRATUSH will employ a short length of optical cable for electrical isolation between the digital and analog receiver. This places high demands on EMI shielding as detailed in Section \ref{sec:EMI}. Alternatively, a hermetically sealed optical transmitter-receiver pair is also being explored along with other RF based reverse isolation options.

\subsubsection{In-situ measurement of antenna return loss}\label{sec:VNA}
PRATUSH employs an in-situ measurement of the complex antenna impedance by use of a custom one-port vector network analyzer (VNA). The VNA functionality is implemented using a dedicated RF and digital section, with in-built calibration and switching schemes designed from first principles. The VNA so realised is optimized for operating over $55-110$ MHz and is sensitive to the expected antenna $S_{11}$ parameter and detect any changes therein.   The reflection coefficient and thereby the transfer function of the antenna depends on the antenna impedance. The in-situ measurement of the complex antenna impedance is motivated by two factors. While the receiver electronics are operated in a thermally regulated environment provided by the satellite package, the antenna radiator and reflector are exposed to multiple thermal cycles over the course of the mission. Despite thermal baffling efforts that will be considered, there could be changes in the antenna electro-mechanical properties as a result of the large thermal swings as the satellite transits over the terminator line (the dividing line between day and night on the Moon), twice per orbit. Besides, an antenna in lunar orbit is immersed in the interplanetary plasma, the characteristics of which is time variable. An antenna in such a plasma may experience impedance changes, as have been observed in wire antennas \citep{WINDWAVES}. The thermal as well as plasma effects will be modeled, and thermal effects quantified in pre-flight tests. These results compounded by in-situ measurements of antenna impedance by the VNA will be used in offline data-analysis to estimate the total antenna efficiency . As shown in Figure~\ref{fig:antenna_validation}, the correction obtained by in-situ measurement is essential to control systematic residuals below mK levels .

\begin{figure}
    \centering
    \includegraphics[width=\textwidth]{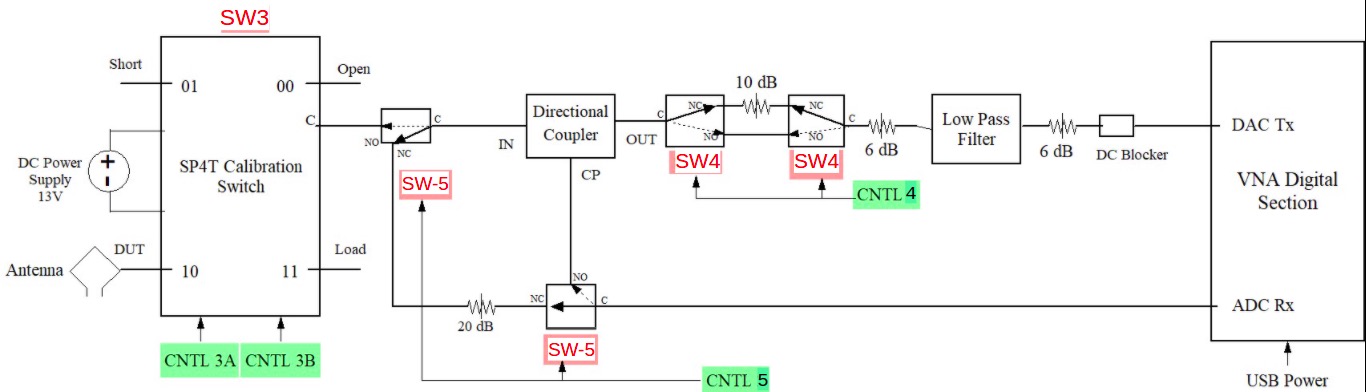}
    \caption{A block diagram representation of the VNA Section in the PRATUSH RF system. The VNA is implemented from first principles using digital and RF electronics to measure the PRATUSH antenna $S_{11}$. The VNA is calibrated using precision terminations namely Open, Short, and $50\Omega$ load.}
    \label{fig:VNA_section}
\end{figure}

The VNA is implemented using a dedicated RF and digital section. A digital-to-analog converter (DAC) in the VNA digital section is used to generate and transmit a tone that sweeps across the frequency range of interest in discrete frequency intervals. The frequency intervals are chosen to most closely match the bin centers of the final sky spectrum as would be recorded by the FPGA based spectrometer when observing the sky. The synthesized signal is looped back for measurement by the analog-to-digital converter (ADC) through a directional coupler to provide the reference power level for further calibration. Following this, the signal from the DAC is then swept over precision termination loads, namely a short, open, and $50~\Omega$ mechanical load. These are introduced in the signal path between the DAC and ADC using three ports of a four port RF switch. The fourth port is connected to the Device Under Test (DUT) whose complex impedance is to be measured, namely the PRATUSH antenna. By using complex (amplitude and phase) measurements from the precision calibrators we derive corrections for the cables, switches, and other RF components in the VNA secion \citep{2016ITMTT..64.2631M}. This correction is in turn applied to the DUT measurement as a calibration of the VNA section, to derive the $S_{11}$ of the PRATUSH antenna. The block diagram of the VNA section alone is shown in Figure~\ref{fig:VNA_section} and the table of operations for VNA calibration as described above is shown in Table.\ref{tab:VNA_measurement}. The equations applied in VNA calibration, implementation and validation details of the VNA will be presented in a future paper.

\begin{table}[]
    \centering
 \resizebox{\textwidth}{!}{
    \begin{tabular}{cccccc}
         Master & SW4 & SW5 & SW3 MSB & SW3 LSB & Measurement \\
         \hline
         ON & 0 & 0 & \multicolumn{2}{c}{\text{Not Applicable}} & Reference signal\\
         ON & 0 & 1 & 0 & 0 & Open \\
         ON & 0 & 1 & 0 & 1 & Short \\
         ON & 0 & 1 & 1 & 0 & Antenna \\
         ON & 1 & 1 & 1 & 1 & 50$\Omega$\\
    \end{tabular}
    }
    \caption{The measurement cycle of the VNA section in the PRATUSH RF system. The master switch connects the antenna to the VNA section. Various control signals are toggled resulting in one of 5 states in which spectra are recorded. These are then used to calibrate the VNA and derive the impedance matching of the PRATUSH antenna. }
    \label{tab:VNA_measurement}
\end{table}

\subsection{PRATUSH Digital electronics}
The digital receiver of PRATUSH performs the following functions.
\begin{itemize}
    \item Digitize the analog input (sky signal) from the antenna via the analog receiver.
    \item Perform windowing of the digital samples 
    \item Realize a FX correlation spectrometer on an FPGA.
    \item Provide control signals for state switching of the analog receiver for calibration.
    \item Process scientific data to reduce data volume suitable to downlink data rates.
    \item Record and store payload scientific data in a single board computer (SBC) and associated memory.
    \item Communicate with the satellite bus electronics for handshaking, control, clocking, and data transmission.
\end{itemize}
The overall architecture and design rules of the digital receiver are based on those implemented for SARAS as presented in \citep{Girish2020}, we summarise the planned implementation strategy for PRATUSH below. For wideband applications in radio astronomy, FPGAs, due to their in-built specialized hardware blocks are well suited to demultiplex high-speed serial data into multiple parallel data streams, enabling a parallelized channelization structure to transform the samples of a wideband time-domain signal into multiple narrow sub-bands for further processing. 
The digital correlation receiver for PRATUSH will comprise two ADCs, a Field Programmable Gate Array (FPGA), associated electronics, data storage, and interface to baseband data handling unit. At the core of the digital receiver would be a real-time digital signal processing platform built around two 12-bit ADCs (one dual ADC) and a space-grade Virtex-5 FPGA. This platform along with an SBC forms the digital backend receiver for PRATUSH. The SBC in addition to processing the data from the FPGA, also acts as the master controller. 

\begin{figure*}
    \centering
    \includegraphics[scale = 0.75, width=\textwidth]{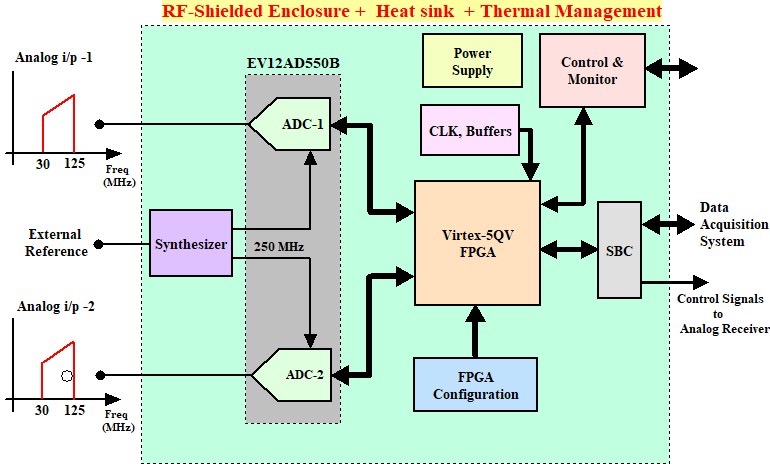}
    \caption{A block diagram representation of the PRATUSH digital receiver. The main components of the PRATUSH digital receiver are a dual channel 12-bit ADC, a Virtex-5 space qualified FPGA, and a space qualified SBC.}
    \label{fig:Digital_BD}
    \end{figure*}

\begin{figure*}
    \centering
    \includegraphics[scale = 0.75, width=\textwidth]{ 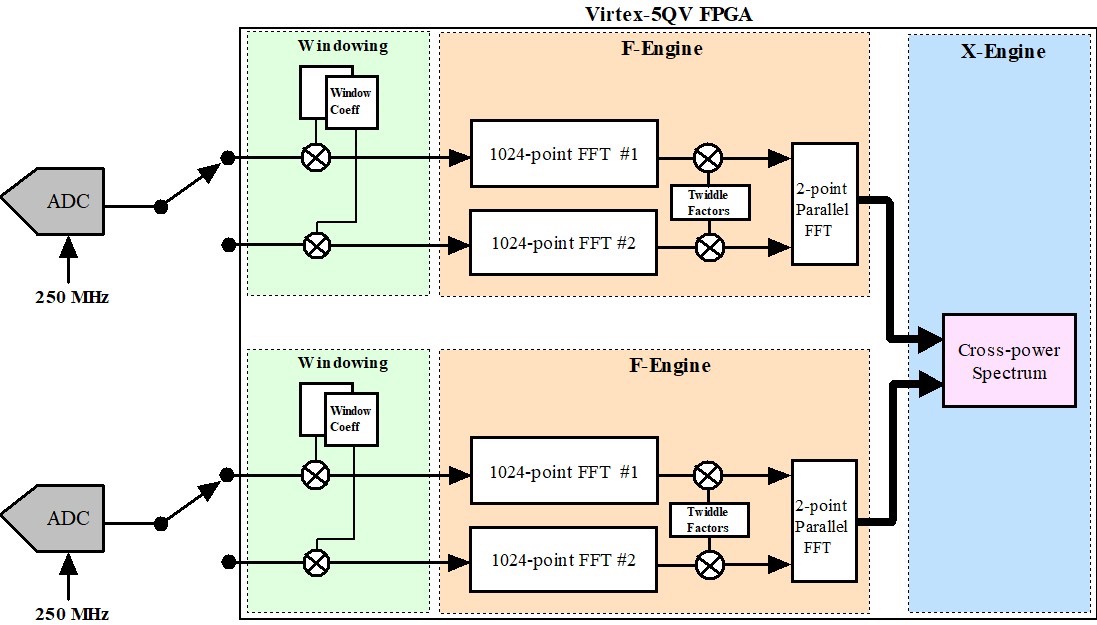}
    \caption{The FPGA firmware architecture implemented in the PRATUSH digital receiver. The digitized signals are weighted with 4 term Nuttall window coefficients, Fourier transformed and multiplied to generate FX averaged cross-power spectrum.} 
    \label{fig:digital_arch}
\end{figure*}

The block diagram of the PRATUSH digital receiver and the FPGA firmware architecture are presented in Figure~\ref{fig:Digital_BD} and \ref{fig:digital_arch} respectively. Two analog signals, in the frequency range 55-110 MHz (3 dB bandwidth), are sampled and digitized using 12-bit ADCs operating at a sampling rate of 250 MHz (4 ns). Virtex 5 FPGA will compute a 2048 point Fast Fourier Transform (FFT) of the buffered data samples streaming out of both ADCs and produce an on-chip averaged (16384 spectra are averaged inside FPGA over a period 134 ms) cross-correlation spectrum of the signals in the two receiver arms. The analog input to the digital receiver unit is cycled through six switching states as described in Table \ref{tab:RF_analog_calibration} above. Hence, the digital receiver needs to operate in Start-Stop mode. This requires a control signal from a Programmable Logic Controller (PLC) to start the digital system. Some parameters that need to be monitored for checking the health of the digital system include voltages of power supplies, status flags inside the FPGA for monitoring of control signals, checksum verification, and total power values.  Status of these parameters needs to be stored in the on-board memory in multiple status information bits of 8-bits each.  The total size of the status information or metadata is estimated to be 2 kilobits.  Inside the Virtex 5 FPGA, the correlation spectrometer will be realized in four distinct stages: grabbing of ADC data samples, weighting of data samples (recommended window function is a 4 word long Nuttall window), Fourier transformation using a 2048-point FFT engine (F-engine) and lastly a multiply-and-accumulate stage (X-engine). The FFT-engine will be implemented as a MxN point architecture where M is the demultiplex factor and N is the length of the FFT. Data from individual ADCs needs to be demultiplexed inside the FPGA to allow multiple paths to operate in parallel at reduced clock speed. In the case of PRATUSH digital receiver, where data streaming from each ADC is at 250 MSps, a demultiplexing factor of two reduces the clock speed of each path to a comfortable 125 MHz. Such a clock speed is within the realm of Virtex-5 FPGA.  

A summary of the specifications of the digital receiver is presented in Table \ref{tab:digital_rx}

\begin{table}[h]
    \centering
     \resizebox{\textwidth}{!}{
    \begin{tabular}{cc }
    Specification & Value \\\\
        Number of analog input & 2 \\
        20 dB Analog signal bandwidth & 30-125 MHz \\
        Sampling speed & 250 MSps \\
        ADC & EV12AD550B\\ 
        ADC bit resolution & 12 bits\\ 
        FPGA & Virtex-5QV (XQR5VFX130)\\ 
        Window function & 4-term Nuttall\citep{Nuttall1981} \\
        Length of FFT & 2048-point \\
        Spectral resolution following windowed FFT & 244 kHz \\
        Output & Averaged cross-correlation spectrum \\
        On-chip integration & 134 ms \\
        Operating temperature range & 0-35 deg Celsius \\
        On-board memory & 8 MB \\
        Power & 35 W \\
    \end{tabular}
    }
    \caption{Specifications of the baseline design PRATUSH digital receiver}
    \label{tab:digital_rx}
\end{table}

\subsection{Comparison with SARAS}
As noted above, SARAS (from the same team) is the mature ground-based counterpart to PRATUSH. Several design choices in PRATUSH are motivated by the experience with SARAS. For instance the bandpass calibration scheme followed in PRATUSH is similar to SARAS-3. So is the requirement of smooth behaviour of sub-systems including the frequency depenendent antenna properties. However, PRATUSH is not a `space-qualified' SARAS experiment.
While the design philosophy of PRATUSH is similar to SARAS, there are some noteable differences which we list below. 
\begin{itemize}
    \item The frequency of operation of PRATUSH baseline design is 55 - 110 MHz. This encompasses the FM frequency band of 88-108 MHz. This frequency range enables maximizing science goal returns (where the spectral richness of the signal from the cosmic dawn is the maximum) in a band that is most affected by terrestrial RFI, and hence presently outside the purview of SARAS. The frequency of the experiment determines all aspects of the antenna and receiver design.
    \item SARAS3 employs a 100 metre long optical fibre cable to electrically and physically isolate the sensitive analog electronics near the antenna from digital electronics placed at a distance. This is not the case in PRATUSH where the antenna sits right above the electronics (at $\sim 0$ metres).
    \item The required shielding effectiveness of the SARAS digital electronics enclosure is $\gtrsim 70$ dB. For PRATUSH this requirement is $\gtrsim 120$ dB factoring in the physical proximity of the electronics to the antenna. This number will further be refined from measurements of the EMI from spacecraft electronics.
    \item SARAS necessarily implements Radio Frequency over Fibre (RFoF) technology to isolate the electronics housed near the antenna from the digital electronics placed over 100 m away, communicating the signal over single-mode optical fibre. PRATUSH will explore and adopt suitable isolation methods based on performance from engineering model tests and technology readiness levels for space use.
    \item The PRATUSH RF-system employs a dedicated VNA section in addition to the analog receiver for bandpass calibration. This has not been implemented in SARAS.
    \item While the digital receiver architecture remains the same between SARAS and PRATUSH, the details of the implementation differ. This comes by replacing the components used with high technology readiness level (TRL) space-qualified counterparts and programming for high reliability, low data volume requirements. For instance the laptop master controller in SARAS is swapped by an SBC, the Virtex-6 FPGA is replaced nominally by a Virtex-5QV radiation-hardened space-grade FPGA (XQR5VFX130), and the EV10AQ190 quad 10-bit ADCs are replaced by a dual channel, space-grade 12-bit ADC (EV12AD550B) in the PRATUSH baseline design.  

\end{itemize}

\subsection{Comparison with other lunar CD/EoR experiments}
PRATUSH will orbit the Moon, preferentially at low altitudes, to maximize the time spent in the far side. Such an observation strategy will provide a recurring science observation window, at the same time alleviating the coupling of antenna with the ground, which is an issue for ground-based experiments \citep{spinelli2022}. However, this necessitates a longer mission duration as only a fraction of time each orbit is available for conducting cosmological observations. Lunar lander missions, such as LuSEE-Night \citep{bale2023lusee}, can conduct uninterrupted observations from the lunar surface. Such observations, while being sensitive to lunar ground profile will be made continuously in a relatively stable environment. The continuous observation can also provide additional methods for foreground separation \cite{2023MNRAS.520..850A}. Both approaches, namely of orbiters and landers have individual and complementary advantages and challenges, making the choice of strategy an interesting and complex risk optimization problem.

\section{Operations}
\begin{figure*}
         \centering
        \includegraphics[scale=0.2]{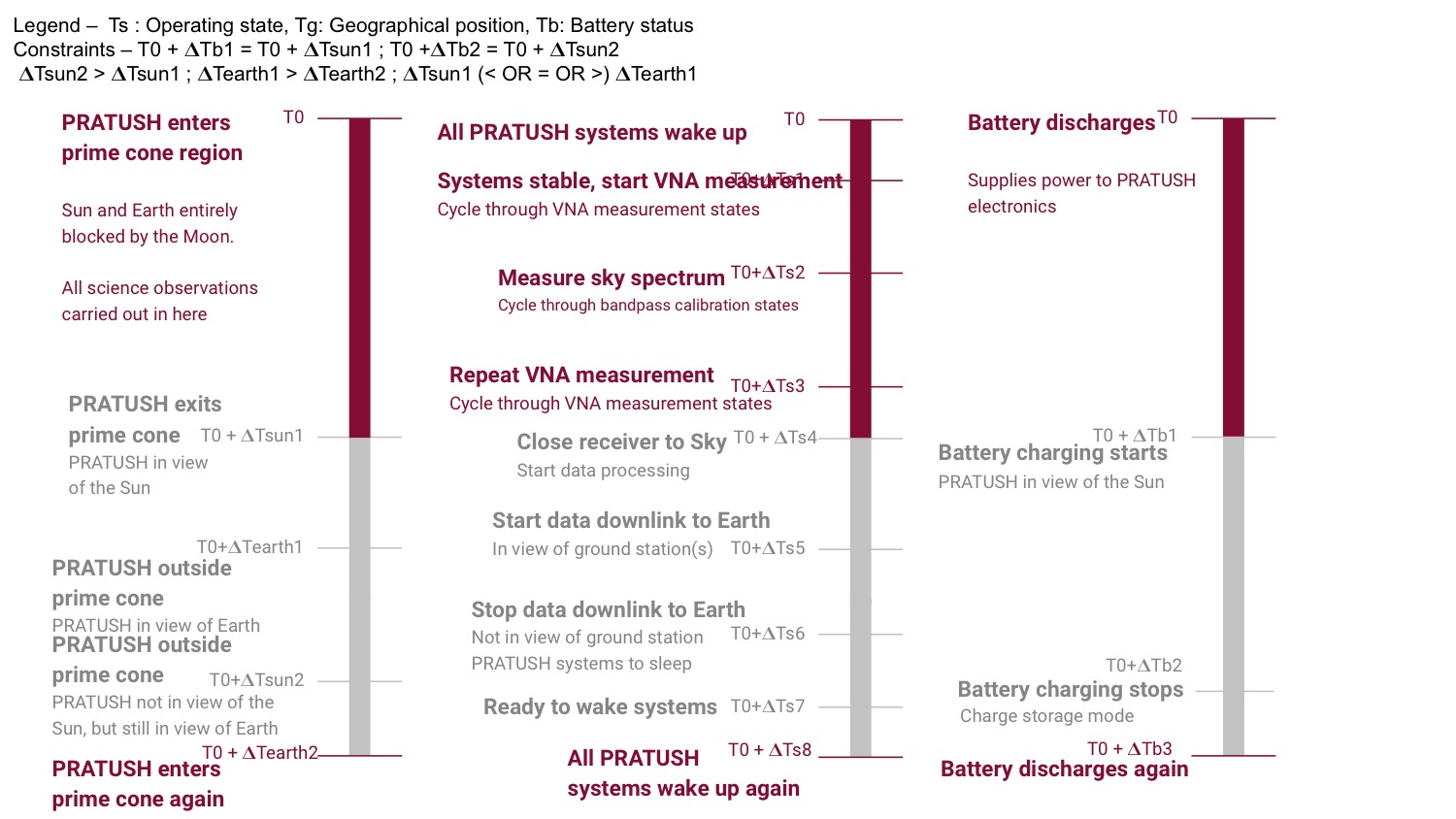}
         \caption{A sample of PRATUSH operations in lunar orbit. Scientific observations are made in the prime cone region. Data is downlinked when in sight of a ground station on Earth. Batteries are charged when the solar panels are illuminated by the Sun.}
        \label{fig:operations}
     \end{figure*}

\begin{figure*}
         \centering
        \includegraphics[scale=0.15]{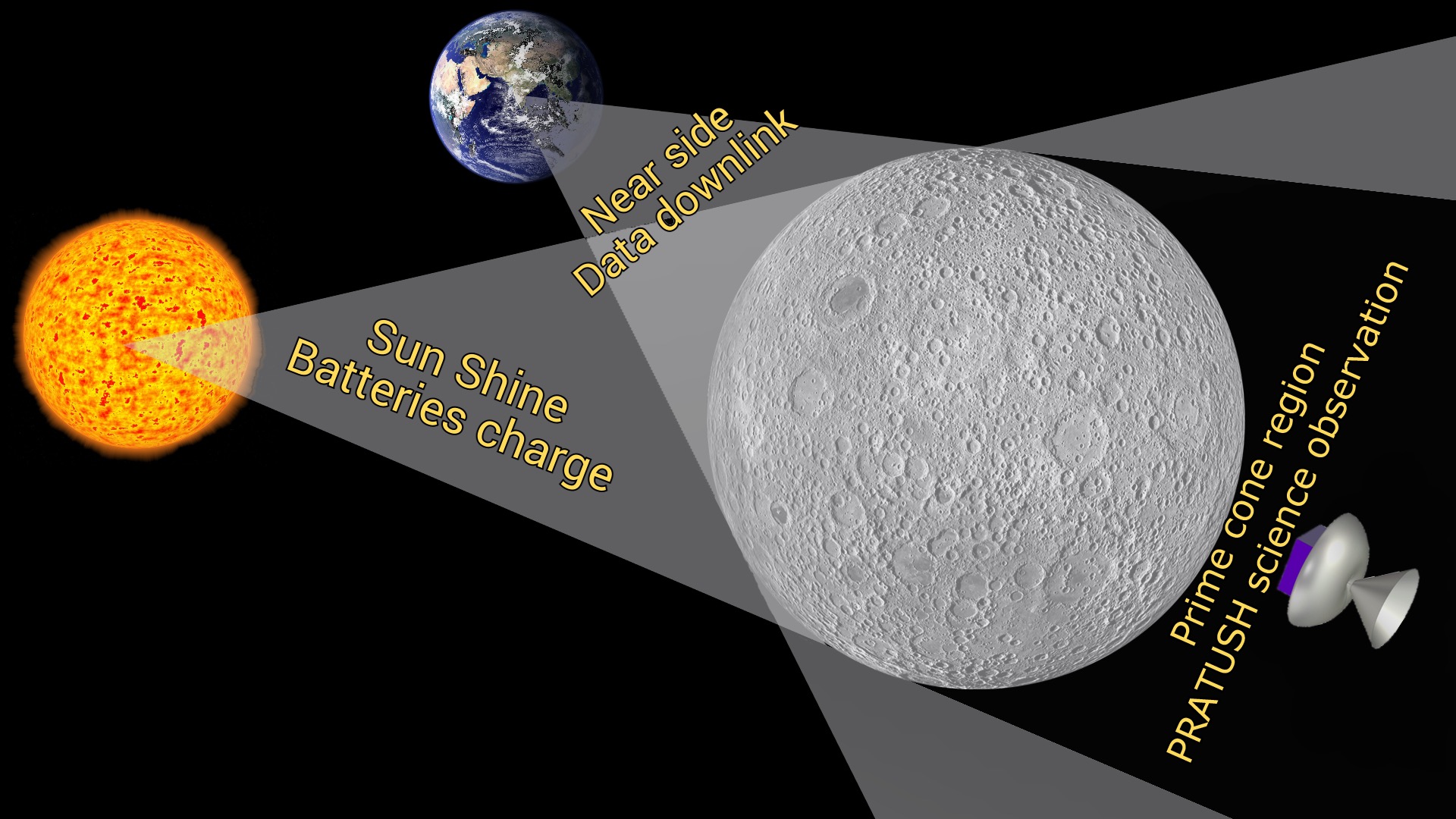}
         \caption{A not-to-scale graphical view of PRATUSH making scientific observations in the lunar farside, when in the prime cone region (indicated by the dark background behind the moon). When illuminated by the sun the solar panels charge the batteries, and when in sight of the Earth, the data is down-linked.}
        \label{fig:PRATUSH_EMS_graphic}
     \end{figure*}

When in orbit around the Moon, PRATUSH will conduct scientific measurements in the prime cone region. The data collected therein will be processed and will be when in sight of ground stations be downlinked. The solar panels housed on the sides of the craft are illuminated when outside the prime cone region and illuminated by the Sun. The orbit determines the specific sequence of operations, a sample of which is demonstrated in Figure~\ref{fig:operations}.
For purposes of PRATUSH operations we consider the position of the spacecraft (PRATUSH) with respect to the three-body system comprising the Moon, the Sun, and the Earth. When PRATUSH enters the prime cone region the payload electronics previously in sleep mode start operations, powered by batteries. Following thermal stabilization of electronics, the antenna return loss is measured by the VNA following the appropriate steps of VNA operation. This is then followed by measurement of the sky-spectrum following the bandpass calibration cycle described in Section.\ref{sec:bpcal}. As PRATUSH nears the end of the prime-cone region, the antenna return loss is once again measured by the VNA and payload electronics enter sleep mode. Depending on the nature of exit from the prime cone region, that is, due to being in sight of the Sun or Earth, different operations are undertaken. When the Sun is visible, the batteries are charged by solar panels illuminated by solar radiation. When the Earth (ground-station) is in the field of view, the spacecraft electronics transmits the data recorded in the previous prime-cone observing cycles, along with meta-data including instrument health. Conversely, as the craft moves out of the view of the Sun and the Earth, battery and data down-linking stop respectively.  As PRATUSH once again enters the prime-cone region the cycle of operations is repeated. A simplified graphic to visualize PRATUSH operations is shown in Figure \ref{fig:PRATUSH_EMS_graphic}.

\section{PRATUSH sensitivity}

\begin{figure*}
         \centering
        \includegraphics[scale=0.35, width=\textwidth]{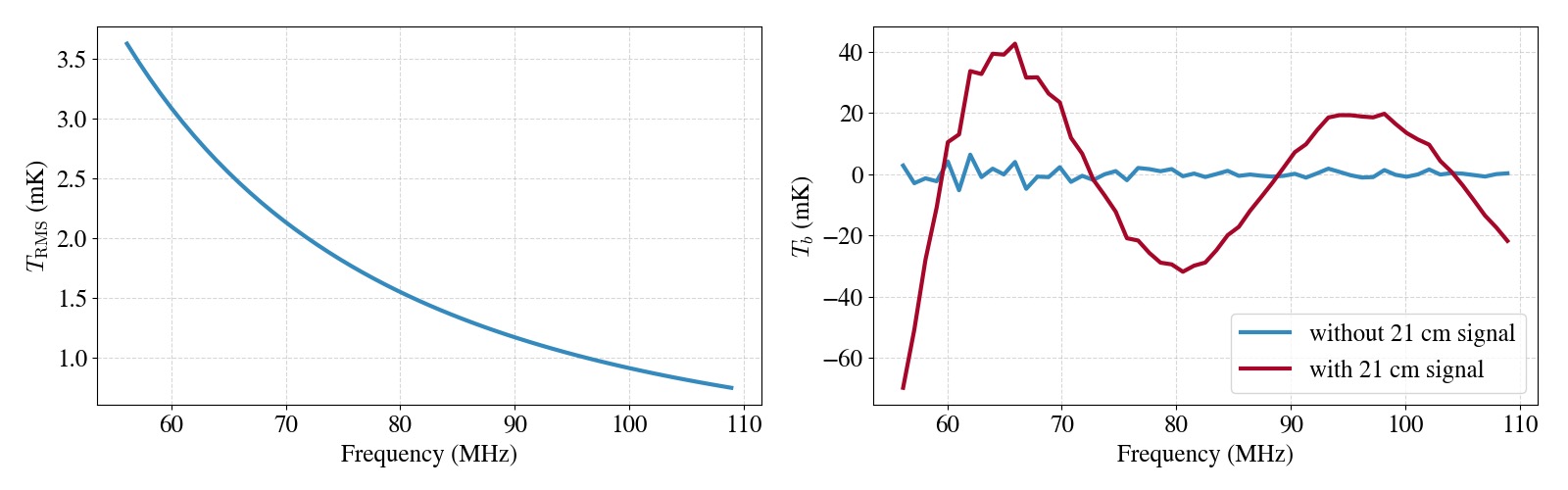}
         \caption{Left panel shows thermal noise r.m.s. over complete mission. Right panel shows residuals after subtracting MS function from the simulated data, with and without adding 21-cm signal in the simulated data.}
        \label{fig:sensitivity}
     \end{figure*}

As noted previously we assume a conservative estimate of 200 hours in the prime-cone region with PRATUSH for a mission lifetime of 2 years. With the current calibration scheme, the observation time will result in $\sim$mK thermal noise r.m.s. at 244~kHz frequency resolution. Figure~\ref{fig:sensitivity} shows the expected thermal noise r.m.s. with frequency on completion of the mission. 

\begin{figure}
         \centering
          \resizebox{\hsize}{!}{\includegraphics{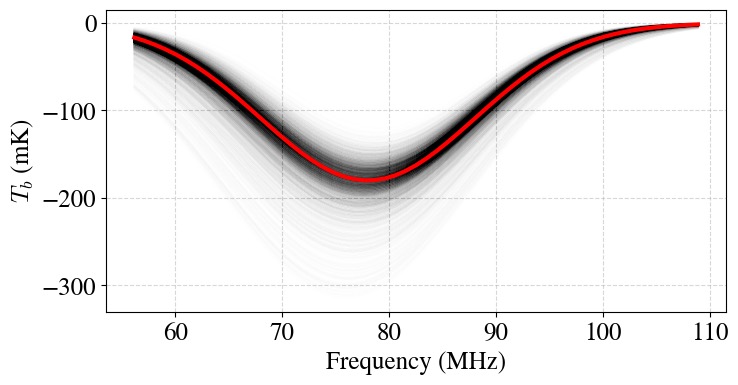}}
         \caption{In red the fiducial signal parameterised as a Gaussian and in black the 21-cm signals generated from sampling the posterior.}
        \label{fig:signal_detection}
     \end{figure}

To assess if this level of sensitivity allows detection of 21-cm signal, we construct mock dataset that comprises GMOSS description of the radio foregrounds, convolved with PRATUSH antenna primary beam, and thermal noise corresponding to the observation time in the prime cone. In one case, we also add 21-cm signal, modeled as a Gaussian with amplitude of -180~mK, centered at 78~MHz, with a full width at half maximum of 23.5~MHz representing a fiducial 21-cm signal as shown by the bright red line in Figure~\ref{fig:signal_detection}, while in another case we do not added any 21-cm signal.
Mock data corresponding to both the cases is fit with a foreground-only model represented by an MS function. Figure~\ref{fig:sensitivity} shows the residuals corresponding to the datasets with and without 21-cm signal. There is a significant increase in the r.m.s. of the residual when the signal is added to the mock data, demonstrating the instrument sensitivity to detect mK amplitude signatures. 

Next, we model the foreground component along with 21-cm signal using 3-parameter Gaussian. Equation~\ref{eq:model} describes the model:

\begin{equation}
    T_\mathrm{model} = \sum_{i=0}^{i=N} c_i \mathrm{log_{10}}(\nu)^i + A~\rm exp{\frac{(\nu - \nu_0)^2}{2 \sigma^2}}.
    \label{eq:model}
\end{equation}

\begin{figure}
     \centering
          \resizebox{\hsize}{!}{\includegraphics{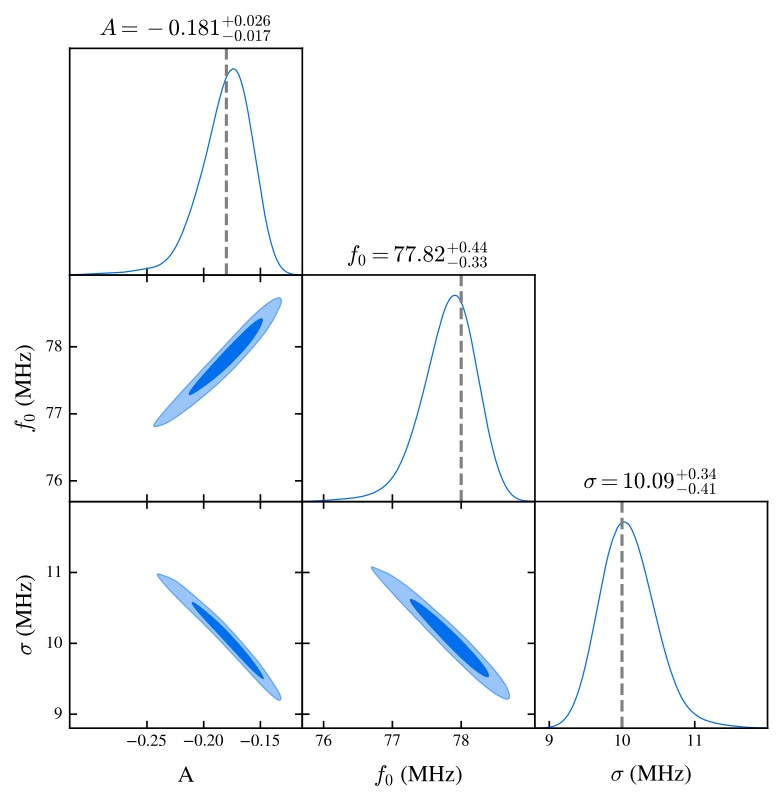}}
         \caption{Posterior on amplitude, centre frequency and width of the Gaussian representing 21-cm signal.}
         \label{fig:gauss_param}
     \end{figure}

We use \texttt{polychord} to optimize the parameters, and extract the 21-cm signal \citep{10.1093/mnras/stv1911}. Figure~\ref{fig:gauss_param} shows the posterior of the 21-cm signal compared to the true parameters and the reconstructed signals respectively in the black shaded region of Figure~\ref{fig:signal_detection}. The sensitivity of PRATUSH, therefore, allows an unbiased estimate of the signal parameters enabling the extraction of 21-cm signal from the foregrounds with high fidelity. 

\section{Current status}
With the PRATUSH baseline design in place, a ground-based concept model is under development. The concept model is expected to lead to the engineering model followed by the flight model subject to experiment (mission) approval. The multi-fold goals of the concept model in leading up to the engineering model are listed below. 
\begin{itemize}
    \item To demonstrate reduction in power-mass-volume specifications of electronics compared to the ground-based predecessor experiment - SARAS \citep{JN2021}.
    \item Expose early risks and challenges in electro-mechanical design of the proposed arrangement with the antenna placed over the craft structure containing the electronics.
    \item To develop confidence in modeling of the instrument to match measurements from the concept model on the ground to enable high fidelity forward modeling of the engineering and flight models. 
    \item Build a preferred parts list and identify critical components with low technology readiness levels to focus developmental activities for space application.
    \item Identify critical paths of EMI leakage in-situ to guide EMI shielding strategies to be adopted in the flight model.
    \item Characterize system performance measures, including receiver noise,  optimize times spent in various calibration states for maximum signal-to-noise ratio.
\end{itemize}

The concept model utilizes a combination of commercially available and custom designed electronics. The final concept model hardware, design, implementation, integration and test results will be presented in a follow-up paper.  


\section{Design considerations}
We highlight here the design challenges that must be considered and addressed for any CD/EoR signal detection experiment in lunar orbit, and hence for PRATUSH as well. These are research and development items that are being actively addressed in the pre-project studies phase.

\subsection{Broadband frequency independent antenna design}
As has been noted previously, designing a frequency independent antenna is a challenging task. In the context of CD/EoR detection experiments, the specific meaning of a frequency independent antenna is one that does not introduce spurious spectral structures, either via mode-coupling of spatial fluctuations to frequency or by non-smooth antenna return loss. While this requirement applies equally to ground-based experiments, the presence of the satellite structure, and potentially metallic structures from other payloads in the near field of the antenna introduces added complexity. In the case of electrical contact, such as via the ground, these otherwise disparate metallic structures become an extension of the CD/EoR detection antenna, introducing asymmetries in the beam and shapes in the return loss. Although the presence of `external' metallic structures cannot be completely avoided, it is essential to design the CD/EoR detection antenna around these metal structures to ensure that the final antenna so arrived at is frequency independent over the full band of operation.

\subsection{Galvanic isolation between RF systems and digital receiver}
It is common practise in sensitive ground-based CD/EoR experiments to use RFoF to provide strong reverse isolation of the RF signal propagating back from the digital receiver towards the RF systems near the antenna. By using separate batteries to power the RF systems and the digital receivers, the two are also galvanically isolated. However, in space, irrespective of the use of the reverse isolation strategies, all power is derived from a common power supply unit via solar panels, and isolating the ground currents of the RF system and Digital receiver is not straightforward. It is thus important to work through the grounding diagram of the full system including the satellite electronics, such that any additional noise due to mixing of noisy and clean grounds is quantified and controlled.

\subsection{Electromagnetic Interference (EMI)} \label{sec:EMI}
Perhaps the biggest technical challenge faced by a global-21 cm experiment in lunar orbit is achieving high levels of suppression of EMI. The purpose of observing the sky in the radio quiet environment offered by the lunar farside is defeated if the payload and spacecraft electronics contaminate measurements in the frequency range of observations. 
 The level of the shielding necessary is given by equation \ref{eq:EMI_SE}.
\begin{equation}\label{eq:EMI_SE}
    SE(\nu, d) \ge Signal(\nu) - EMI(\nu,d)\  [dB],
\end{equation}
wherein, $SE$ is the shielding effectiveness at distance $d$, which needs to exceed the difference between the $Signal$ strength and $EMI$ strength over the full frequency range $\nu$.
We start with identifying the mechanisms by which EMI can contaminate the measured spectrum. Broadly these are via conductive and radiative modes. Radiated EMI is most pernicious when picked up by the antenna, which is the primary sensing element of the scientific payload. EMI that enters the receiver via the antenna is not removed by the bandpass calibration, as it is an additive to the sky-signal. EMI enclosures of high shielding effectiveness are aimed at containing or limiting the contaminant signal strength that is measured outside the enclosure.
As the name suggests, conductive EMI is that which enters the signal path via cables, as an additive contaminant to the sky-signal. This is to be distinguished from additives that result from impedance mismatch within the receiver electronics. 
EMI suppression must be achieved using a multi-pronged strategy. We suggest a few below. While this is not an exhaustive list, a suitable combination of these strategies adapted to the specific instrument can alleviate the detrimental effects of EMI. 
\begin{itemize}
    \item Multi-layer EMI shielding : Electronics are to be packaged in individual modular chassis each offering a layer of EMI shielding. In addition to shielding of cards and modules, entire sub-systems namely of analog receiver and digital receiver electronics are to be further packaged in shielding enclosures. The design of EMI shielding enclosures is a rich subject with appropriate attention to detail around access points using gaskets, shielded connectors for signal and power, thermal management. Furthermore choice of materials plays an important role in achieving high shielding effectiveness while providing lightweight options to reduce launch costs. 
    \item Choice of clocking frequencies: The dominant sources of EMI are expected to be clocks in the digital electronics of the experiment and space-craft. Though expected to be localized to fixed frequencies, the effect of these strong tones can degrade performance over multiple channels in the observation band by way of harmonics and intermodulation distortions. A careful choice of the clocking frequency and clock source is advisable such that the strong tone lies at a channel center in the measured spectrum. This allows for a clean removal of the tone by excising the channel where the tone lies and a few surrounding channels by means of suitable windowing functions and offline flagging. A related and important tool to ensure minimal spectral leakage of RFI across channels is the use of the ``picket-fence" algorithm as an FFT window as described in \cite{pulupa2017}.
    \item Linear power supplies : While clocks are expected to generate localized EMI that appear as strong spikes within or between channels, and their harmonics, switched mode power supplies are known to generate broadband EMI. Broadband EMI can result in a change in the shape of the instrument bandpass, increase the noise floor resulting in systematics that can confuse or limit signal detection. Thus an appropriate choice of linear power supplies that can supply the required voltages are a better choice for power distribution.
    \item Good EMI shielding practices : There are rules of thumb of good EMI shielding practice that are advisable in any experiment. To name a few, these include use of twisted pair of wires or coaxial cables where appropriate, adopting of a sandwiched routing scheme of signal layer between two shielding layers in PCB design (in particular of flexible PCBs), use of space qualified ferrite beads around cables and connectors. 
    \item Rigorous on-ground testing : On-ground tests of the effective measured spectrum in the presence of powered on and operational experiment electronics as well as space-craft electronics is vital to arriving at the final in-flight EMI shielding strategy. Any remnant low-level tones in localised channels measured with high confidence on ground can provide a template for cross-referencing and subtraction of on-sky data transmitted from orbit.
    \item Distance : A standard practise to reduce the level of EMI received by the antenna from the receiver electronics is by increasing the distance between the two. The Friis transmission formula suggests a loss of $6~dB$ in RF power for every doubling of distance between a transmiter and receiver in free-space. With this in mind it is common practise in ground-based experiments to have a separation of as much as 100 metres or longer between the antenna element and the digital receiver of the telescope, where every doubling of distance affords $\sim 6 dB$ of extra shielding \citep{trainotti1990}. In space it may not be feasible to achieve such large distances. Since power levels reduce as the square of distance as per Friis transmission formula, any separation between the digital receiver and spacecraft electronics, such as by means of a boom would be advantageous from an EMI shielding perspective.
\end{itemize}

\subsection{On satellite data processing}
Ground-based global CD/EoR signal detection experiments have been quick to adopt the rapid advances in data transfer technologies. These include connecting data-acquisition and processing electronics to computers or servers via dedicated high-speed gigabit Ethernet and optical interfaces. The high speed connectivity enables large volumes of raw data transfer and data acquisition at high cadence, enabling thorough RFI excision and processing on analysis servers. Current high TRL data downlink technologies limit the downlink speeds for space-based CD/EoR signal detection experiments. There is precedent by experimental demonstration of transmitting at 622 Mbps using free space optical communication from Moon to Earth with the Lunar Laser Communication Demonstration \citep{LADEE_demo} in NASA's Lunar Atmosphere and Dust Environment Explorer (LADEE) mission \citep{LADEE}. This is encouraging to achieve continued high data downlink speeds for future missions. Current standard technology enabled by radio wave communication limits bandwidths to achieve data downlink rates of $\sim 100~$ Mbps \citep{LRO}. This has consequences to algorithm implementation, calibration cadence, data throughput and, averaging times. A mission specific trade off must be achieved between on-satellite processing requirements (and hence power, mass, volume) and the ability to analyse data in its most raw format on the ground. The high dynamic range of the problem also necessitates double precision computing to avoid introducing truncation or round off related systematic errors, which sets requirements on the choice of space-qualified SBCs for on-satellite data processing to reduce data volumes conducive to downlinking.  

\subsection{Reflection of Galactic emission off lunar regolith}
A wide antenna beam that is characteristic of CD/EoR experiments is expected to receive radiation from the Moon as a fraction of the total emission observed. The emission from the Moon can be broadly divided into two components namely the intrinsic brightness temperature of the Moon and the galactic radio emission reflected off the lunar surface. The former maybe modeled as a blackbody of temperature of $\sim 200$ Kelvin, which behaves as a frequency-independent or constant additive to the total observed sky spectrum  \citep{baldwin1961}. The latter however is a complex function determined by the properties of the lunar regolith, its skin depth and resulting frequency dependent effects. Efforts to simulate these effects will provide an understanding of the expected effect on CD/EoR signal detection. If foregrounds and their spectral behavior are well constrained, a sensitive lunar orbiter like PRATUSH would itself provide the data products which may be used to reconstruct properties of the lunar regolith, the challenge being disentangling frequency structures from regolith induced chromaticity and the CD/EoR signal. 

\section{Orbital requirements}
We list below the criteria for deciding the orbital parameters of a lunar orbiter experiment for detecting the CD/EoR signal in the lunar farside. The primary metric of the orbit is the effective science observation time. This is the time spent in the shadow region of the Earth and the Sun, behind the Moon. This ensures minimal terrestrial RFI as well as attenuation of high radio flux levels emitted by transient events arising from solar activity, and provides a stable and consistent environment for observations. With the time spent in the lunar farside as the driving metric, secondary technological factors are to be optimized and determined for an effective science mission, as detailed below.

\subsection{Optimizing for thermal cycles}
The orbiter enters the lunar farside from the nearside through a transitory region crossing over the terminator line. This results in large thermal swings, especially on external subsystems that are not temperature regulated. The antenna, which is the primary sensing element is most likely to experience mechanical and thermal stress over multiple orbits and consequently multiple thermal cycles. Mechanical changes to the antenna shape can significantly change the antenna properties of return loss (impedance matching) and beam pattern. These effects must be simulated and tested on ground in pre-flight studies. Thus, lunar orbiters carrying CD/EoR detection payloads sensitive to variations in antenna electro-mechanical properties need a combination of in-situ antenna characterization and thermal regulation by way of baffles. 

\subsection{Effective science quality data duration}
Simulations have demonstrated that in ideal conditions where the system temperature is foreground dominated and noise is purely thermal, the CD/EoR signal can be detected within minutes of on-sky integration \citep{MSR2017b}. However realizing such a measurement is harder in practise due to  instrument generated systematics, necessitating multiple hours of on-sky observations. The primary factor that guides orbital parameter selection is the effective amount of time that the payload spends in the lunar farside observing the sky. In the prime cone regin, the CD/EoR detection experiment will spend a fraction of the time on calibrator sources (such as internal noise reference) which is time away from the sky. Time spent for thermal stabilization of electronics, on internal calibrators, and time lost to spectral contamination from satellites, lunar landers, or other rogue sources of RFI also need to be deducted from time in the Sun-Earth shadow to provide the effective time gathering science quality data. 

An expectation of the total science quality observing time determined by target sensitivity and cost driving technology such as propellant for orbit and altitude correction determine the total mission lifetime. Conversely, the expected sensitivity of the experimental payload is determined by the total mission lifetime and usable on-sky observation time, determining the types of signals and detection thresholds that the  experiment is sensitive to.

\subsection{Pointing accuracy and stabilization}
Most ground-based global CD/EoR detection experiments operate in a sky drift mode. Unlike traditional steerable (including digital steering) radio telescopes, global CD/EoR telescopes have large beams and do not use bright compact astrophysical objects for flux or phase calibration \footnote{Note that global CD/EoR experiments do use sky models, such as to derive antenna efficiency \cite{SARAS3}.}. Deployed in a remote relatively radio-quiet location, the antenna beam is directed skywards and the receiver collects data in the form of sky-spectra. These are timestamped by GPS conditioned receivers. The antenna beam pattern and the LST derived using site location determines the sky region observed at any instant and hence the scan strategy. A similar strategy can be adopted in lunar orbit in space.  With global signal detection experiments employing wide antenna beams (such as half power beam width $\sim 60^{\circ}$), a Position, Navigation and Timing (PNT) system along with a star tracker can provide a positional accuracy of $\sim 1^{\circ}$. In addition to orbit and altitude correction to maintain the specific orbit requirements of the payload, three-axis spin stabilization is required for pointing the PRATUSH antenna away from the Moon. \\\\
Significant work has been done by the DARE team in identifying a ``frozen orbit" around the moon requiring minimum correction for orbital stability for a mission duration of 2 years \cite{dare_mission_design}. This can be investigated incorporating the latest models of lunar gravity as a potential orbit for PRATUSH.

\section{Summary}
PRATUSH is a proposed lunar orbiter experiment with the primary science goal of detecting the redshifted global 21-cm signal from CD/EoR in the radio quiet environs of the lunar farside. The baseline design for PRATUSH has been presented in this paper. The payload design leverages experience from the legacy ground-based experiment SARAS \citep{Patra2015,SS2018a,JN2021}. While the principles for a CD/EoR detection experiment remain the same between ground and space measurements, there are some design considerations unique to a space-based lunar orbiter CD/EoR detection experiment. These are summarized as an experiment design checklist with proposed solutions that maybe customized to any specific experiment with a primary science goal of detecting the faint, global, 21-cm signal from the period of formation of the very first stars and galaxies, and the subsequent reionization of the Universe.

Though space-based experiments are inherently more complex and expensive than comparable ground-based experiments, the two are to be seen as complementary to one another to make a clear, high confidence cosmological signal detection. The faint and poorly constrained nature of the standard model signals from CD and EoR (let alone the exotic models) makes the detection extremely challenging, as evidenced by decades of efforts spent on the signal detection from the ground. The confidence in detection claimed will likely be countered by challenge of the host of unknown-unknowns that cannot be modeled and hence accounted for. This will remain the case for lunar farside experiments alone. However, a detection of the same signal from experiments operating in very different environments namely on Earth and in space, with likely very different systematics, will be the highest indicator of cosmological origin of any signal detection reported. Thus, farside lunar experiments are not to be seen as an alternative to, but as an essential complement to ground-based detection experiments. As ground-based experiments are already approaching the regime of making detection claims, constraining and ruling out models, the time is now ripe to plan and fly lunar farside experiments. This will ensure that these lunar farside experiments are ready with results in time for the upcoming data release from the ongoing and upcoming ground based EoR telescopes.  While experiments on ground battle with terrestrial RFI, the lunar farside environment is believed to remain pristine. However, with several upcoming lunar missions, including commercial flights, orbiters, and a host of activity (including scientific instruments), this might not remain the case for long.

While there has been a recommendation to preserve the lunar farside as a protected radio-quiet zone by the Moon Farside Protection Permanent Committee, this has not yet been ratified by international policy. Therefore, it is essential to make observations in the lunar farside to take advantage of the pristine environment it offers fo for CD/EoR  detection in a timely manner.

\bibliography{mybibfile}
\vspace{2cm}
\textbf{Funding declaration}\\
The authors acknowledge the Indian Space Research Organisation (ISRO) for funding the work presented in this paper under grant SSPO:Astronomy-AO:2019/-07.\\\\

\textbf{Author contributions}\\
MSR led the writing of the manuscript.
MSR and SS1 led the full system design presented in the paper. SS1 led the work and writing of section 5.2.1. KSS and BSG led the digital subsystem design and the writing of section 5.3 along with contributions from AA. KS and RS led the RF system design including section 5.2 along with contributions VG. AR and KK led the antenna subsystem design, namely section 5.1. NUS and SS2 provided overall suggestions through the work and editorial advise of the manuscript.\\\\

\textbf{Data Availability}\\
Data sets generated during the current study are available from the corresponding author on reasonable request.\\\\

\textbf{Conflict of Interest}\\
The authors have no relevant financial or non-financial interests to disclose.

\end{document}